\documentclass[prd,floatfix,amsmath,amssymb,nofootinbib,twocolumn,reprint]{revtex4-1}
\usepackage{graphicx,color,dcolumn,booktabs,bm,adjustbox,esint,upgreek}
\usepackage{longtable,comment,braket}
\usepackage{amsfonts,mathrsfs,amsmath}
\usepackage{times}
\usepackage{overpic}
\usepackage{tipa}
\usepackage{indentfirst}
\usepackage{feynmf}   
\usepackage{slashed}  
\usepackage{cases}
\usepackage{epstopdf}
\usepackage{psfrag}
\usepackage{subfigure}
\usepackage{capt-of}
\usepackage{dcolumn,kantlipsum}
\usepackage{color}
\usepackage{multirow}
\usepackage{ulem}
\usepackage{mathtools}
\usepackage[colorlinks, citecolor=blue,anchorcolor=red,menucolor=red,linkcolor=blue,filecolor=red,runcolor=red,urlcolor=blue,frenchlinks=red]{hyperref}

\allowdisplaybreaks[4]

\begin{document}
\title{How to understand the $X(2900)$?}
\author{Bo Wang}\email{wangbo@hbu.edu.cn}
\affiliation{School of Physical Science and Technology, Hebei University, Baoding 071002, China}
\affiliation{Key Laboratory of High-precision Computation and Application of Quantum Field Theory of Hebei Province, Baoding 071002, China}

\author{Shi-Lin Zhu}\email{zhusl@pku.edu.cn}
\affiliation{School of Physics and Center of High Energy Physics, Peking University, Beijing 100871, China}

\begin{abstract}
In this work, the $S$- and $P$-wave $\bar{D}^\ast K^\ast$
interactions are studied in a coupled-channel formalism to
understand the recently observed $X_0(2900)$ and $X_1(2900)$ at
LHCb. The experimental event distributions can be well described,
and two states with $I(J^P)=0(0^+)$ and $0(1^-)$ are yielded in a
unified framework. The masses of the $0^+$
and $1^-$ states are consistent with the experimental data, but the
width of the $0^+$ state is larger than that of the $1^-$ one. The
$X_1(2900)$ can be interpreted as the $P$-wave excitation of the
ground-state $X_0(2900)$ in the hadronic molecular picture. The $S$-
and $P$-wave multiplets in the $\bar{D}^\ast K^\ast$ system have
many members, so the present peak in the $D^-K^+$ invariant mass
distributions might contain multi substructures.
\end{abstract}

\maketitle

\section{Introduction}\label{Introduction}
Recently, the LHCb Collaboration observed a clear peak in the
$D^-K^+$ invariant mass spectrum of the $B^+\to D^+D^-K^+$
decay~\cite{LHCb:2020bls,LHCb:2020pxc}, where the helicity angle
distribution shows evident $P$-wave behavior. The peak is fitted
with spin-$0$ and spin-$1$ states. Their resonance parameters are
determined to be
\begin{eqnarray}
X_0(2900): M&=&2866\pm7\pm2~\text{MeV},\nonumber\\
\Gamma&=&57\pm12\pm4~\text{MeV};\nonumber\\
X_1(2900): M&=&2904\pm5\pm1~\text{MeV},\nonumber\\
\Gamma&=&110\pm11\pm4~\text{MeV}.\nonumber
\end{eqnarray}
The $D^-K^+$ decay channel implies their quark components should be
$\bar{c}\bar{s}ud$, which means that they are the fully open flavor
exotic hadrons.

Many theoretical interpretations have been proposed to understand
the inner structures of $X(2900)$, such as the molecular states from
the $\bar{D}^\ast K^\ast$ and $\bar{D}_1K$
interactions~\cite{Chen:2020aos,He:2020btl,Liu:2020nil,Hu:2020mxp,Agaev:2020nrc},
the compact tetraquarks
$\bar{c}\bar{s}ud$~\cite{Chen:2020aos,Karliner:2020vsi,He:2020jna,Wang:2020xyc,Zhang:2020oze,Wang:2020prk},
and kinematic effects from the triangle
singularities~\cite{Liu:2020orv,Burns:2020epm}. The production and
decay properties were investigated in
Refs.~\cite{Huang:2020ptc,Chen:2020eyu,Burns:2020xne,Xiao:2020ltm}.
One can also consult
Refs.~\cite{Albuquerque:2020ugi,Lu:2020qmp,Mutuk:2020igv,Tan:2020cpu,Abreu:2020ony,Qi:2021iyv,Chen:2021erj,Hsiao:2021tyq,Duan:2021bna,Kong:2021ohg,Dong:2020rgs,Bondar:2020eoa,Chen:2021tad}
for other pertinent works.

In Ref.~\cite{Karliner:2020vsi}, the $X_0(2900)$ was interpreted as
the $S$-wave compact tetraquark in a string-junction picture, while
the subsequent calculations in Ref.~\cite{He:2020jna} supported it
to be the radial excited tetraquark with $J^P=0^+$. Meanwhile, the
obtained masses with the refined quark model calculations in
Refs.~\cite{Wang:2020prk,Lu:2020qmp,Tan:2020cpu} are much lower than
the measured mass of $X_0(2900)$. In other words, the isosinglet
$S$-wave compact tetraquark $\bar{c}\bar{s}ud$ can not reconcile
with the $X_0(2900)$. A typical feature of $X_0(2900)$ is its mass
below the $\bar{D}^\ast K^\ast$ threshold about $30$ MeV. The QCD
sum rule calculations~\cite{Chen:2020aos,Agaev:2020nrc} and the
one-boson-exchange inspired models~\cite{He:2020btl,Liu:2020nil} all
supported the $S$-wave $\bar{D}^\ast K^\ast$ molecule for
$X_0(2900)$. Besides, Refs.~\cite{Huang:2020ptc,Xiao:2020ltm}
investigated the decays of $X_0(2900)$ and indicated large
$\bar{D}^\ast K^\ast$ component in its wave function. Another hint
comes from the width of $X_0(2900)$, which is close to the $K^\ast$
width, i.e., one may infer that the $X_0(2900)$ has an intrinsic
$K^\ast$ width, and the residual part arises from its decays. Thus,
it seems the $X_0(2900)$ is more likely to be the $\bar{D}^\ast
K^\ast$ molecular state.

Most of the previous works cannot give a unified description for
$X_0(2900)$ and $X_1(2900)$ if they are the genuine states. For
example, Ref.~\cite{Karliner:2020vsi} relegated the $X_1(2900)$ to
the $\bar{D}^\ast K^\ast$ rescattering effect.
Ref.~\cite{He:2020btl} gave a virtual state explanation for
$X_1(2900)$ that was generated from the $\bar{D}_1K$ interaction and
the bound solution in the $\bar{D}_1K$ channel needs an unnaturally
large cutoff. In Ref.~\cite{Liu:2020nil}, the $X_1(2900)$ cannot be
interpreted as the $P$-wave $\bar{D}^\ast K^\ast$ molecule. If the
$X_1(2900)$ is indeed the $P$-wave compact tetraquark as suggested
in Refs.~\cite{Chen:2020aos,He:2020jna}, then where is the $S$-wave
ground state? It definitely cannot be the $X_0(2900)$ as the
results shown in
Refs.~\cite{He:2020jna,Wang:2020prk,Lu:2020qmp,Tan:2020cpu}. In
addition, we also need to answer the different dynamics in
$b$-decays for the formations of  $X_0(2900)$ and $X_1(2900)$. In
order to eschew the dilemmas as mentioned above, this work is
devoted to describing these two states in a unified framework, and
understanding their internal configurations in a more natural way.

With the discoveries of more and more near-threshold exotic
states~\cite{Chen:2016qju,Guo:2017jvc,Liu:2019zoy,Lebed:2016hpi,Esposito:2016noz,Brambilla:2019esw,Chen:2021ftn,Chen:2022asf,Meng:2022ozq},
we find the connections between hadronic physics and nuclear physics
become closer and closer. The concepts of
$\chi$EFT~\cite{Weinberg:1990rz,Weinberg:1991um}, which have been
successfully substantialized to describe the nuclear
forces~\cite{Bernard:1995dp,Epelbaum:2008ga,Machleidt:2011zz,Meissner:2015wva,Hammer:2019poc,RodriguezEntem:2020jgp}, can also be
generalized to depict the interactions between heavy
hadrons~\cite{Meng:2022ozq,AlFiky:2005jd,Fleming:2007rp,Baru:2011rs,Valderrama:2012jv,Braaten:2015tga,Schmidt:2018vvl,Wang:2018atz,Meng:2019ilv,Wang:2019ato,Wang:2020dko,Meng:2020cbk,Chen:2021htr}.
This work dedicates to model the
$\bar{D}^\ast K^\ast$ interactions with $S$- and $P$-waves in a
coupled-channel formalism via mimicking the manipulations in $\chi$EFT. By fitting the $D^-K^+$ event
distributions in experiments, we analyze the pole positions in the
scattering $T$-matrix to see whether the $X_0(2900)$ and $X_1(2900)$
can be assigned as the $S$-wave and $P$-wave excited $\bar{D}^\ast
K^\ast$ molecules. This can definitely resolve the problem we are
facing now, and give a uniform description of the $X_0(2900)$ and
$X_1(2900)$.

This paper is organized as follows. In Sec.~\ref{EFT}, we establish
the effective potentials of $\bar{D}^\ast K^\ast$ and iterate them
into the coupled-channel Lippmann-Schwinger equations. In
Sec.~\ref{NumAndDis}, we present our numerical results and
discussions. In Sec.~\ref{Sum}, we conclude this work with a short
summary. 

\section{$\bar{D}^\ast K^\ast$ rescattering in coupled-channel formalism}\label{EFT}

In this work, we assume the $X_0(2900)$ and $X_1(2900)$ are the
isosinglet $S$- and $P$-wave hadronic molecules which are generated
from the $\bar{D}^\ast K^\ast$ interactions, respectively. Thus the
flavor wave functions in $\bar{D}^\ast K^\ast$ and $\bar{D}K$
channels can be written as
\begin{eqnarray}
|\bar{D}^\ast K^\ast,I=0\rangle&=&\frac{1}{\sqrt{2}}[\bar{D}^{\ast0}K^{\ast0}-D^{\ast-}K^{\ast+}],\nonumber\\
|\bar{D}
K,I=0\rangle&=&\frac{1}{\sqrt{2}}[\bar{D}^{0}K^{0}-D^{-}K^{+}].
\end{eqnarray}

The mass of $K^\ast$ is close to the nucleon mass, thus $K^\ast$ can
be regarded as a heavy matter field. We can generalize the
approaches of $\chi$EFT as that in the $NN$ system~\cite{Bernard:1995dp,Epelbaum:2008ga,Machleidt:2011zz,Meissner:2015wva,Hammer:2019poc,RodriguezEntem:2020jgp} to the
$\bar{D}^\ast K^\ast$ case. The effective potentials in the
$\bar{D}^\ast K^\ast$ system can be parameterized  as follows,
\begin{eqnarray}\label{Vtotal}
\mathcal{V}=\sum_i V_i(\bm p^\prime,\bm p)\mathcal{O}_i(\bm
p^\prime,\bm
p,\bm{\varepsilon},\bm{\varepsilon}^\dagger,\bm{\varepsilon}^\prime,\bm{\varepsilon}^{\prime\dagger}),
\end{eqnarray}
where $\bm p$ and $\bm p^\prime$ represent the three momenta of the
initial and final states in the center of mass system (c.m.s) of
$\bar{D}^\ast K^\ast$, respectively. $\bm{\varepsilon}^{(\prime)}$
and $\bm{\varepsilon}^{(\prime)^\dagger}$ denote the polarization
vectors of initial and final $\bar{D}^\ast$ ($K^\ast$),
respectively. $V_i$ are the scalar functions of $\bm p$ and $\bm
p^\prime$. $\mathcal{O}_i$ are the pertinent operators that can be
constructed from the scalar products among unit operator $\bm 1$,
vectors $\bm p^{(\prime)}$, $\bm{\varepsilon}^{(\prime)}$ and
$\bm{\varepsilon}^{(\prime)^\dagger}$, which read
\begin{eqnarray}\label{Operators}
\mathcal{O}_1&=&(\bm\varepsilon^{\dagger}\cdot\bm\varepsilon) (\bm\varepsilon^{\prime\dagger}\cdot\bm\varepsilon^{\prime}),\nonumber\\
\mathcal{O}_2&=&(\bm\varepsilon^{\prime\dagger}\cdot\bm\varepsilon)(\bm\varepsilon^\dagger\cdot\bm\varepsilon^\prime)-(\bm\varepsilon^{\prime\dagger}\cdot\bm\varepsilon^\dagger)(\bm\varepsilon\cdot\bm\varepsilon^\prime),\nonumber\\
\mathcal{O}_3&=&(\bm q\cdot\bm \varepsilon^{\prime\dagger})(\bm q\cdot\bm\varepsilon)(\bm\varepsilon^\dagger\cdot\bm\varepsilon^\prime)+(\bm q\cdot\bm\varepsilon^\dagger)(\bm q\cdot\bm\varepsilon^\prime)(\bm\varepsilon^{\prime\dagger}\cdot\bm\varepsilon),\nonumber\\
\mathcal{O}_4&=&(\bm q\cdot \bm\varepsilon^\dagger)(\bm q\cdot\bm\varepsilon^{\prime\dagger})(\bm\varepsilon^\prime\cdot\bm\varepsilon)+(\bm q\cdot\bm\varepsilon)(\bm q\cdot\bm\varepsilon^\prime)(\bm\varepsilon^{\prime\dagger}\cdot\bm\varepsilon^\dagger),\nonumber\\
\mathcal{O}_5&=&(\bm k\cdot\bm \varepsilon^{\prime\dagger})(\bm k\cdot\bm\varepsilon)(\bm\varepsilon^\dagger\cdot\bm\varepsilon^\prime)+(\bm k\cdot\bm\varepsilon^\dagger)(\bm k\cdot\bm\varepsilon^\prime)(\bm\varepsilon^{\prime\dagger}\cdot\bm\varepsilon),\nonumber\\
\mathcal{O}_6&=&(\bm k\cdot \bm\varepsilon^\dagger)(\bm k\cdot\bm\varepsilon^{\prime\dagger})(\bm\varepsilon^\prime\cdot\bm\varepsilon)+(\bm k\cdot\bm\varepsilon)(\bm k\cdot\bm\varepsilon^\prime)(\bm\varepsilon^{\prime\dagger}\cdot\bm\varepsilon^\dagger),\nonumber\\
\mathcal{O}_7&=&(\bm\varepsilon^{\prime\dagger}\cdot\bm\varepsilon^\prime)(\bm\varepsilon^{\dagger}\times\bm\varepsilon)\cdot(\bm k\times\bm q)\nonumber\\
&&+(\bm\varepsilon^{\dagger}\cdot\bm\varepsilon)(\bm\varepsilon^{\prime\dagger}\times\bm\varepsilon^\prime)\cdot(\bm
k\times\bm q),\dots,
\end{eqnarray}
with $\bm q=\bm p-\bm p^\prime$ the transferred momentum and $\bm
k=(\bm p^\prime+\bm p)/2$ the average momentum. The ellipsis denotes the higher terms with more $\bm q$ and $\bm k$. The operators $\mathcal{O}_1$, $\mathcal{O}_2$, $\mathcal{O}_{3,\dots,6}$ and $\mathcal{O}_7$  account for the central force, the spin-spin interaction, the tensor force, and the spin-orbital ($SL$) coupling, respectively.


In order to describe the transitions between $\bar{D}^\ast K^\ast$
and $\bar{D}K$, we define the corresponding transition operators,
\begin{eqnarray}\label{OperatorsIn}
\mathcal{O}_{1}^\prime&=&\bm\varepsilon\cdot\bm\varepsilon^\prime\text{~or~}\bm\varepsilon^\dagger\cdot\bm\varepsilon^{\prime\dagger},\nonumber\\
\mathcal{O}_{2}^\prime&=&(\bm q\cdot\bm\varepsilon)(\bm
q\cdot\bm\varepsilon^\prime)\text{~or~}(\bm
q\cdot\bm\varepsilon^\dagger)(\bm
q\cdot\bm\varepsilon^{\prime\dagger}),\nonumber\\
\mathcal{O}_{3}^\prime&=&(\bm k\cdot\bm\varepsilon)(\bm
k\cdot\bm\varepsilon^\prime)\text{~or~}(\bm
k\cdot\bm\varepsilon^\dagger)(\bm
k\cdot\bm\varepsilon^{\prime\dagger}).
\end{eqnarray}
For performing the partial wave decompositions in the following, the polarization vectors in above equations can be transformed into the spin transition operator $\bm S_t$ with the approach given in the appendix C of Ref.~\cite{Wang:2019ato}.

The effective potentials of the elastic ($\bar{D}^\ast K^\ast\to\bar{D}^\ast K^\ast$) and the inelastic ($\bar{D}^\ast
K^\ast\to\bar{D} K$, or in reverse) channels have both contributions from the contact and one-pion-exchange (OPE) interactions. The short-range contact terms for the elastic (el) and inelastic (in) scatterings are given as, respectively,
\begin{eqnarray}
\mathcal{V}_{\text{ct}}^{\text{el}}&=&\sum_{i=1}^{7}C_i\mathcal{O}_i+(C_{8}\bm q^2+C_{9}\bm k^2)\mathcal{O}_1\nonumber\\
&&+(C_{10}\bm q^2+C_{11}\bm k^2)\mathcal{O}_2,\\
\mathcal{V}_{\text{ct}}^{\text{in}}&=&C_2^\prime\mathcal{O}_2^\prime+C_3^\prime\mathcal{O}_3^\prime+(C_4^\prime \bm q^2+C_5^\prime\bm k^2)\mathcal{O}_1^\prime,
\end{eqnarray}
where $C_i^{(\prime)}$ are the low energy constants (LECs). 

The long-distance interaction for $\bar{D}^\ast K^\ast$ is depicted
by the OPE contribution. The OPE effective
potential can be easily obtained from the chiral Lagrangians, in
which the anticharmed meson and pion coupling reads~\cite{Wise:1992hn,Manohar:2000dt}
\begin{eqnarray}\label{DpiCoup}
\mathcal{L}_{\tilde{\mathcal{H}}\varphi}=g_{\varphi}\langle\bar{\tilde{\mathcal{H}}}\gamma^\mu\gamma_5
u_\mu\tilde{\mathcal{H}}\rangle,
\end{eqnarray}
where $g_{\varphi}\simeq0.57$ is the axial coupling.
$\tilde{\mathcal{H}}$ and $u_\mu$ are the superfield for anticharmed
mesons and axial-vector current for pion, respectively, their
expressions can be found in Refs.~\cite{Meng:2019ilv,Wang:2019ato}.
In SU(2) case, if we treat the $\bar{s}$ quark as a relatively heavy
quark, then the $K^\ast$ and pion coupling can be formulated as the
same form as that in Eq.~\eqref{DpiCoup}. Its axial coupling
constant $g_\varphi\simeq1.12$ is determined from the decay width of
$K^\ast\to K\pi$~\cite{ParticleDataGroup:2020ssz}.

\begin{figure}[!hptb]
\begin{centering}
    \scalebox{1.0}{\includegraphics[width=0.8\linewidth]{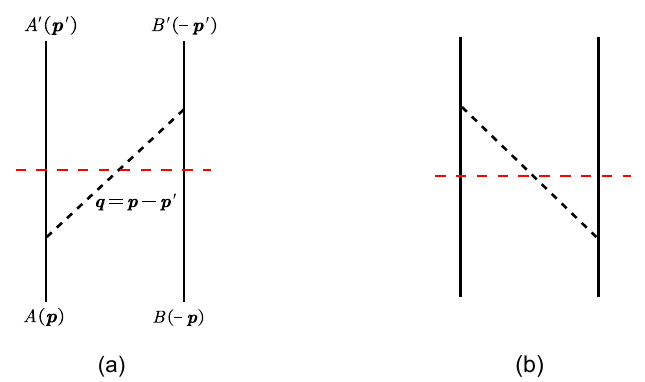}}
    \caption{The OPE diagrams in time-ordered perturbation theory. The black solid and dashed lines denote the $\bar{D}^{(\ast)}/K^{(\ast)}$ and pion, respectively. The red-dashed (horizontal) line indicates the time at which the intermediate state is evaluated. In diagram (a), the
particle $B$ feels the effect of a pion that from the source particle $A$ emitted at an earlier time, thus the $A$ and $B$ particles interact in this case through a retarded propagator. In diagram (b), from the $B$'s point of view, the
effect is felt before the source $A$ emitted the pion, thus the pion propagator in this case is called an advanced propagator.\label{TOPT}}
\end{centering}
\end{figure}

The OPE effective potentials for elastic and inelastic scatterings
read
\begin{eqnarray}
\mathcal{V}_{\text{OPE}}^{\text{(a)}}&=&-\frac{3g_\varphi^2}{4f_\varphi^2}\frac{\mathcal{O}^{\text{ch}}}{2E_\pi[E_\pi+E_{A^\prime}+E_B-E]},\label{OPEel}\\
\mathcal{V}_{\text{OPE}}^{\text{(b)}}&=&-\frac{3g_\varphi^2}{4f_\varphi^2}\frac{\mathcal{O}^{\text{ch}}}{2E_\pi(E_\pi+E_{A}+E_{B^\prime}-E)},\label{OPEin}
\end{eqnarray}
where the $\mathcal{V}_{\text{OPE}}^{\text{(a)}}$ and $\mathcal{V}_{\text{OPE}}^{\text{(b)}}$ represent the contributions from Figs.~\ref{TOPT}(a) and~\ref{TOPT}(b), respectively.
The $\mathcal{O}^{\text{ch}}=-\bm q^2\mathcal{O}_2+\mathcal{O}_3-\mathcal{O}_4$ for the elastic channel, and $\mathcal{O}^{\text{ch}}=\mathcal{O}_{2}^\prime$ for the inelastic channels, respectively. $g_\varphi=\sqrt{0.57\times1.12}$ is the redefined coupling constant,
$f_\varphi=92.4$ MeV is the pion decay constant, and $m_\pi\simeq140$
MeV is the pion mass. Note that the non-static effect (the energy dependence) for the OPE potentials is considered with the time-ordered perturbation theory (see Refs.~\cite{Baru:2011rs,Du:2019pij,Wang:2018jlv,Baru:2019xnh}).
The energies appearing in Eqs.~\eqref{OPEel} and \eqref{OPEin} are given as
\begin{eqnarray}
E_\pi&=&\sqrt{\bm q^2+m_\pi^2},\\
 E_{i^{(\prime)}}&=&m_{i^{(\prime)}}+\frac{p^{(\prime)2}}{2m_{i^{(\prime)}}},\quad i=A,B,
\end{eqnarray}
and $E$ stands for the total energy of the system.

\begin{figure}[!hptb]
\begin{centering}
    \scalebox{1.0}{\includegraphics[width=\linewidth]{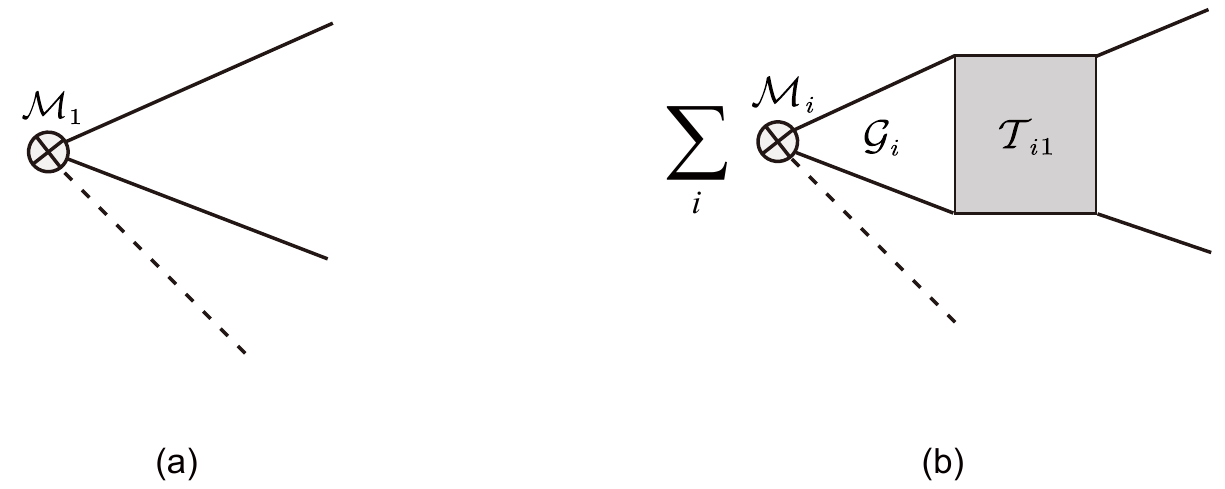}}
    \caption{Diagrams (a) and (b) describe the direct production and rescattering contribution, respectively. The gray circle with cross represents the effective $B^+\to D^+\bar{D}K$ and $B^+\to D^+\bar{D}^{\ast}K^{\ast}$ coupling, while the gray box in diagram (b) signifies the rescattering $T$-matrix of the $\bar{D}^\ast K^\ast$ system. The dashed line denotes the spectator $D^+$, while the solid lines stand for the involved $\bar{D}K$ ($\bar{D}^{\ast}K^{\ast}$) in the rescattering.\label{Production}}
\end{centering}
\end{figure}

The $X_0(2900)$ and $X_1(2900)$ are observed in the $b$-decay
process $B^+\to D^+D^-K^+$, so we need to simulate this production
process. The corresponding Feynman diagrams are shown in
Fig.~\ref{Production}. The reactions in Fig.~\ref{Production} can be
expressed with the coupled-channel Lippmann-Schwinger equations
(LSEs),
\begin{eqnarray}\label{LSEs}
&&\mathcal{U}_\alpha^j(E,\bm p)=\nonumber\\
&&\mathcal{M}_\alpha^j+\sum_\beta\int\frac{d^3\bm
q}{(2\pi)^3}\mathcal{V}_{\alpha\beta}^j(E,\bm p,\bm
q)\mathcal{G}_\beta(E,\bm q)\mathcal{U}_\beta^j(E,\bm q),
\end{eqnarray}
where the subscript $\alpha(\beta)=1,2$ denotes the corresponding
channels [the $\bar{D}K$ and $\bar{D}^\ast K^\ast$ channels are
labeled as $1$ and $2$, respectively, e.g., see
Fig.~\ref{Production}], while the superscript $j$ represents the
fixed total angular momentum. $\mathcal{M}$ is the direct production
amplitude. $E$ stands for the invariant mass of the $D^-K^+$ system.
$\mathcal{G}$ is the Green's function of the intermediate states,
which read
\begin{eqnarray}\label{eq:GreenF}
\mathcal{G}_\beta(E,\bm q)=\frac{2\mu_\beta}{\bm p_\beta^2-\bm
q^2+i\epsilon}, |\bm
p_\beta|=\sqrt{2\mu_\beta(E-m_{\text{th}}^\beta)},
\end{eqnarray}
with $\mu_\beta$ and $m_{\text{th}}^\beta$ the reduced mass and the
threshold of the $\beta$-th channel. The width of $K^\ast$ is $50.8$
MeV~\cite{ParticleDataGroup:2020ssz}, which is considered in
$m_{\text{th}}^\beta$ by using a complex mass $m-i\Gamma/2$ for the
$K^\ast$~\cite{Du:2019pij}.

In order to solve the LSEs in Eq.~\eqref{LSEs}, we need to make the
partial wave decomposition. The effective potentials in
Eq.~\eqref{Vtotal} are given in the plane wave helicity basis, which
can be transformed to the partial wave ($\ell sj$) basis via the
approach in Ref.~\cite{Golak:2009ri}. If the $X_0(2900)$ and
$X_1(2900)$ are the $S$- and $P$-wave $\bar{D}^\ast K^\ast$ molecules
with $J^P=0^+$ and $J^P=1^-$, respectively. Then the corresponding
effective potentials in partial wave bases with the coupled-channels
\begin{eqnarray}
j=0:|\bar{D}K\rangle_{^1S_0},|\bar{D}^\ast
K^\ast\rangle_{^1S_0},|\bar{D}^\ast
K^\ast\rangle_{^5D_0}\label{sbasis}
\end{eqnarray}
and
\begin{eqnarray}
j&=&1:|\bar{D}K\rangle_{^1P_1},|\bar{D}^\ast
K^\ast\rangle_{^1P_1},|\bar{D}^\ast
K^\ast\rangle_{^3P_1},|\bar{D}^\ast
K^\ast\rangle_{^5P_1}\label{pbasis}\nonumber\\
\end{eqnarray}
can be written as the $3\times3$ and $4\times4$ matrices,
respectively.

The contact interactions through the partial
wave decompositions in $S$-wave with the coupled-channels read
\begin{equation}\label{eq:VctS}
[\mathcal{V}_{\text{ct}}]_{\ell,\ell^\prime}^S=\left[
  \begin{array}{ccc}
    0 & \mathcal{C}_{12}(p^2+p^{\prime2}) & \mathcal{C}_{13}p^2\\
    \mathcal{C}_{12}(p^2+p^{\prime2}) & \mathcal{C}_{22}+\mathcal{C}_{22}^\prime(p^2+p^{\prime2}) & \mathcal{C}_{23}p^2\\
    \mathcal{C}_{13}p^{\prime2}&\mathcal{C}_{23}p^{\prime2}&0
  \end{array}
\right],
\end{equation}
where
\begin{eqnarray}
\mathcal{C}_{12}&=&-\frac{\pi}{\sqrt{3}}(4C_2^\prime+C_3^\prime+12C_4^\prime+3C_5^\prime),\nonumber\\
\mathcal{C}_{13}&=&\frac{\sqrt{2}}{\sqrt{3}}\pi(4C_2^\prime+C_3^\prime),\nonumber\\
\mathcal{C}_{22}&=&4\pi(C_1-2C_2),\nonumber\\
\mathcal{C}_{22}^\prime&=&\frac{\pi}{3}(8C_3+24C_4+2C_5+6C_6+12C_8+3C_{9}\nonumber\\
&&-24C_{10}-6C_{11}),\nonumber\\
\mathcal{C}_{23}&=&-\frac{\sqrt{2}\pi}{3}(8 C_3 + 12 C_4 + 2 C_5 + 3 C_6),
\end{eqnarray}
while the ones for the $P$-wave are given as
\begin{equation}\label{eq:VctP}
[\mathcal{V}_{\text{ct}}]_{\ell,\ell^\prime}^P=\left[
  \begin{array}{cccc}
    0 & \mathbb{C}_{12}pp^\prime & 0& \mathbb{C}_{14}pp^\prime\\
    \mathbb{C}_{12}pp^\prime & \mathbb{C}_{22}pp^\prime & 0& \mathbb{C}_{24}pp^\prime\\
    0&0&\mathbb{C}_{33}pp^\prime& 0\\
    \mathbb{C}_{14}pp^\prime&\mathbb{C}_{24}pp^\prime&0& \mathbb{C}_{44}pp^\prime
  \end{array}
\right],
\end{equation}
where
\begin{eqnarray}
\mathbb{C}_{12}&=&\frac{2\pi}{3\sqrt{3}}(4C_2^\prime-C_3^\prime+12C_4^\prime-3C_5^\prime),\nonumber\\
\mathbb{C}_{14}&=&\frac{2\sqrt{5}\pi}{3\sqrt{3}}(4C_2^\prime-C_3^\prime),\nonumber\\
\mathbb{C}_{22}&=&\frac{2\pi}{9}(-8 C_3 - 24 C_4 + 2 C_5 + 6 C_6 - 12 C_8 + 3 C_9\nonumber\\
&&+24 C_{10} - 6 C_{11}),\nonumber\\
\mathbb{C}_{24}&=&\frac{2\pi}{9\sqrt{5}}(-40 C_3 - 60 C_4 + 10 C_5 + 15 C_6 + 12 C_7),\nonumber\\
\mathbb{C}_{33}&=&\frac{\pi}{3}(12 C_3 - 3 C_5 - 8 C_7 -8 C_8 + 2 C_9\nonumber\\
&&+8 C_{10} - 2 C_{11}), \nonumber\\
\mathbb{C}_{44}&=&-\frac{\pi}{45}(220 C_3 - 55 C_5 + 144 C_7 + 120 C_8 \nonumber\\
&&- 30 C_9+120 C_{10} - 30 C_{11}).
\end{eqnarray}
In above potentials, we switch off the $\bar{D}K$ interaction for reducing the free
parameters, which may be described by the chiral
perturbation theory~\cite{Liu:2009uz}. However, the $\bar{D}K$ lies
far below the $\bar{D}^\ast K^\ast$ threshold, thus their
interactions should be irrelevant to the physics around the
$\bar{D}^\ast K^\ast$ threshold.

The LSEs in Eq.~\eqref{LSEs} is regulated with a form factor, in
which the $\mathcal{V}_{\ell,\ell^\prime}$ is multiplied by the
Gaussian regulator,
\begin{eqnarray}\label{eq:gsformf}
\mathcal{V}_{\ell,\ell^\prime}\to
\mathcal{V}_{\ell,\ell^\prime}\exp\left[-\frac{p^2}{\Lambda^2}-\frac{p^{\prime2}}{\Lambda^{\prime2}}\right].
\end{eqnarray}
At the $X(2900)$ energy, the center of mass momentum of the $\bar{D}K$
channel is $0.74$ GeV, and hence we use a relatively hard cutoff
$\Lambda=1.0$ GeV for this channel. For the elastic channel scattering, we vary the
cutoff $\Lambda^\prime$ [in this case, the $\Lambda$ and $\Lambda^\prime$ in Eq.~\eqref{eq:gsformf} have the equal value] over a wide range, i.e., from a soft scale $0.3$ GeV to a hard scale $1.0$ GeV.

Additionally, we also need to mimic the production vertex. We assume
the $D^+D^-K^+$ and $D^+D^{\ast-}K^{\ast+}$ are produced from a
point-like source. Then the $S$- and $P$-wave production amplitudes
can be parametrized as~\cite{Back:2017zqt}
\begin{eqnarray}
&&\mathcal{M}^S_{\{B^+\to D^+D^-K^+,B^+\to D^+D^{\ast-}K^{\ast+}\}_\alpha}\nonumber\\
&&~\quad=\{\tilde{g}_s,\tilde{g}_s^\prime\bm\varepsilon^\dagger\cdot\bm\varepsilon^{\prime\dagger}\},\label{MPs}\\
&&\mathcal{M}^P_{\{B^+\to D^+D^-K^+,B^+\to D^+D^{\ast-}K^{\ast+}\}_\alpha}\nonumber\\
&&~\quad=\{\tilde{g}_p\bm k_1\cdot\bm k_2,\tilde{g}_p^\prime(\bm k_1\cdot\bm
k_2^\prime)(\bm\varepsilon^\dagger\cdot\bm\varepsilon^{\prime\dagger})\nonumber\\
&&~\quad+\tilde{g}_p^{\prime\prime}(\bm k_1\cdot\bm\varepsilon^\dagger)(\bm k_2^\prime\cdot\bm\varepsilon^{\prime\dagger})+\tilde{g}_p^{\prime\prime\prime}(\bm k_1\cdot\bm\varepsilon^{\prime\dagger})(\bm k_2^\prime\cdot\bm\varepsilon^{\dagger})\}\label{MPp},
\end{eqnarray}
where $\tilde{g}_{s(p)}^{(\prime)}$ represents the production
strengths in corresponding channels. $\bm k_1$ and $\bm
k_2^{(\prime)}$ denote the three momentum of $D^+$ and $D^{-}$
($D^{\ast-}$) in the c.m.s of $D^-K^+$ ($D^{\ast-}K^{\ast+}$).

Projecting the production amplitudes to the bases in
Eqs.~\eqref{sbasis} and~\eqref{pbasis}, respectively, one easily
obtains
\begin{eqnarray}
&&\mathcal{M}^{j=0}_{\{B^+\to D^+D^-K^+,B^+\to D^+D^{\ast-}K^{\ast+}\}_i}\label{eq:rs}\nonumber\\
&&~\quad=\mathcal{N}_s\{1,\mathcal{R}_s,0\},\\
&&\mathcal{M}^{j=1}_{\{B^+\to D^+D^-K^+,B^+\to D^+D^{\ast-}K^{\ast+}\}_i}\label{eq:rp}\nonumber\\
&&~\quad=\mathcal{N}_p\{k_1 k_2,\mathcal{R}_p k_1 k_2^\prime,\mathcal{R}_p^\prime k_1 k_2^\prime,\mathcal{R}_p^{\prime\prime} k_1 k_2^\prime\},
\end{eqnarray}
where the irrelevant factors are absorbed into the
$\mathcal{N}_{s(p)}$, while the $\mathcal{R}_{s(p)}$ describes the
relative strengths between $|\bar{D}K\rangle_{^1S_0}$ and
$|\bar{D}^\ast K^\ast\rangle_{^1S_0}$ ($|\bar{D}K\rangle_{^1P_1}$
and $|\bar{D}^\ast K^\ast\rangle_{^1P_1}$) channels, and similar meanings for the $\mathcal{R}_p^{\prime,\prime\prime}$.

With the above preparations, the differential decay width for
$B^+\to D^+D^-K^+$ reads
\begin{eqnarray}\label{eq:rsp}
\frac{d\Gamma}{dE}=\mathcal{N}\frac{|\tilde{\bm k}_1||\bm
k_2^\ast|}{4(2\pi)^3m_B^2}\left[\mathcal{R}_{sp}\left|\mathcal{U}_1^{j=0}\right|^2+\left|\mathcal{U}_1^{j=1}\right|^2\right],
\end{eqnarray}
where an overall normalization factor $\mathcal{N}$ is used to match
the event distributions in experiments. $\tilde{\bm k}_1$ and
$\bm{k}_2^\ast$ are the three momentum of the spectator $D^+$ in the
c.m.s of $B^+$ and the three momentum of $D^{-}$ in the c.m.s of
$D^-K^+$, respectively. $\mathcal{R}_{sp}$ is introduced to
designate the ratio $\mathcal{N}_s^2/\mathcal{N}_p^2$.

\section{Numerical results and discussions}\label{NumAndDis}

Unlike the $D_{s0}(2317)$ and
$D_{s1}(2460)$~\cite{ParticleDataGroup:2020ssz}, there is no
$\bar{c}\bar{s}$ bare core in the hadron spectrum, so the $X(2900)$
provides us a relatively clean environment to study the
coupled-channel effects between open-charm and open-strange mesons.

\begin{figure*}[hptb]
\begin{centering}
    \scalebox{1.0}{\includegraphics[width=\linewidth]{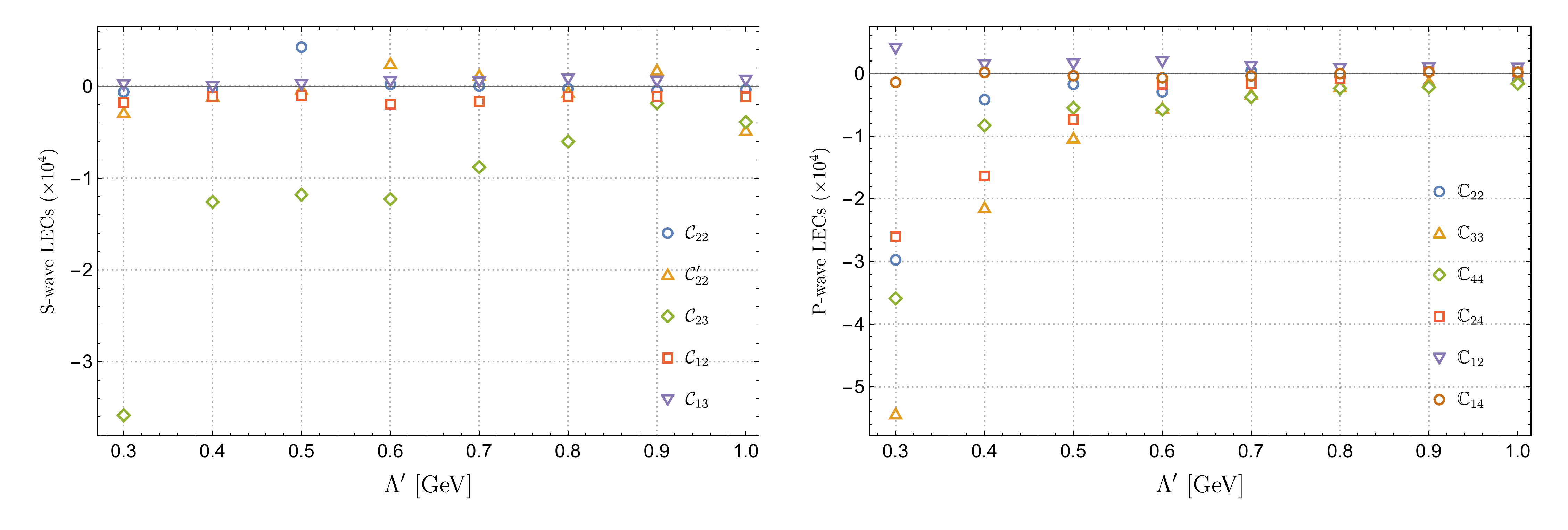}}
    \caption{Changes of the $S$- and $P$-wave LECs when the elastic channel's cutoff is varied from $0.3$ GeV to $1.0$ GeV and the inelastic channel's cutoff is fixed at $1.0$ GeV.\label{LECsPlot}}
\end{centering}
\end{figure*}

The values of partial wave LECs are constrained by
fitting to the $D^-K^+$ event candidates that measured by the
LHCb~\cite{LHCb:2020pxc}, which  are given in Fig.~\ref{LECsPlot} for the $S$- and $P$-waves.
One can find that some LECs are sensitive to the variations of cutoff, such as $\mathcal{C}_{23}$, $\mathbb{C}_{22}$, $\mathbb{C}_{33}$, $\mathbb{C}_{44}$ and $\mathbb{C}_{24}$ [in effective field theory, the LECs are cutoff dependent, $C_i\equiv C_i(\Lambda)$, i.e., the cutoff dependence is absorbed by the LECs, which makes the observables cutoff independent]. The distribution behaviors of these LECs with the cutoff is similar to that of the spin singlet case of $NN$ scattering~\cite{Nogga:2005hy}, while the LECs of the spin triplet $NN$ (the channel for deuteron) scattering show limit-cycle-like behavior (an unsmooth dependence on the cutoff). Meanwhile, the other LECs are not very sensitive to cutoffs, such as the $\mathcal{C}_{12}$, $\mathcal{C}_{13}$,  $\mathbb{C}_{12}$ and  $\mathbb{C}_{14}$. A more general analysis on the nonrelativistic two-body scattering of the single-channel case with renormalization-group invariance constraints was provided by Birse {\it et al}~\cite{Birse:1998dk}.

\begin{figure}[hptb]
\begin{centering}
    \scalebox{1.0}{\includegraphics[width=\linewidth]{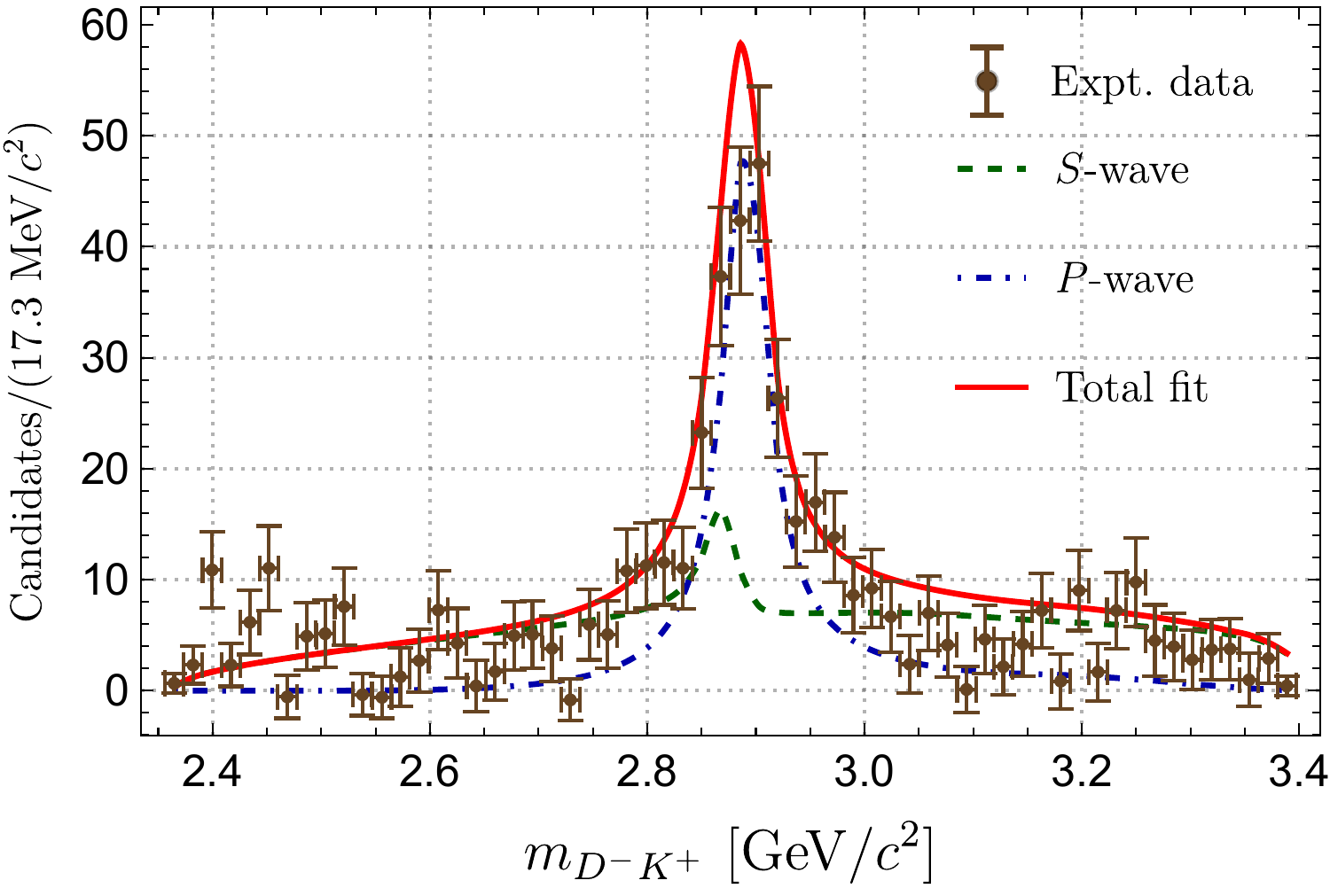}}
    \caption{The fitted $D^-K^+$ invariant mass distributions for $B^+\to D^+D^-K^+$ decay. The experimental data are extracted from Ref.~\cite{LHCb:2020pxc}, where the reflection contributions from the charmonia are subtracted. The dashed, dot-dashed and solid lines denote the $S$-, $P$- and $S+P$-wave contributions, respectively. The lineshapes are obtained with the cutoffs $\Lambda=1.0$ GeV, and $\Lambda^\prime=0.5$ GeV.\label{FitPlot}}
\end{centering}
\end{figure}

The fitted line shape of the $S$- and $P$-wave components, as well as their total contributions are plotted in Fig.~\ref{FitPlot}. We find the results are also insensitive to cutoffs, the $\chi^2/{\text{d.o.f}}$ is around $1.3$ when $\Lambda^\prime\in[0.3,1.0]$ GeV.
The fit with $\Lambda=1.0$ GeV, $\Lambda^\prime=0.5$ GeV gives the least $\chi^2$ ($\simeq1.25$), so we adopt the results in this fit as our outputs. The corresponding values of LECs with the errors are given in Table.~\ref{tab:lecs}.
\begin{table*}
\centering
\renewcommand{\arraystretch}{1.5}
\caption{The values of LECs $(\mathcal{C}_{22},\mathcal{C}_{22}^{\prime},\mathcal{C}_{23},\mathcal{C}_{12},\mathcal{C}_{13},\mathbb{C}_{22},\mathbb{C}_{33},\mathbb{C}_{44},\mathbb{C}_{24},\mathbb{C}_{12},\mathbb{C}_{14})/10^4$ with errors that fitted at $\Lambda=1.0$ GeV and $\Lambda^\prime=0.5$ GeV.\label{tab:lecs}}
\setlength{\tabcolsep}{3.3mm}
{
\begin{tabular}{cccccc}
  \hline
 $\mathcal{C}_{22}$ (GeV$^{-2}$)&	$\mathcal{C}_{22}^{\prime}$ (GeV$^{-4}$)& 	$\mathcal{C}_{23}$ (GeV$^{-4}$)&	$\mathcal{C}_{12}$ (GeV$^{-4}$)&	$\mathcal{C}_{13}$ (GeV$^{-4}$)\\
$0.428\pm0.0823$&	$-0.0466\pm0.0151$&	$-1.18\pm0.324$&	$-0.102\pm0.0435$&	$0.0336\pm0.0117	$\\
\hline
$\mathbb{C}_{22}$ (GeV$^{-4}$)&	$\mathbb{C}_{33}$ (GeV$^{-4}$)&	$\mathbb{C}_{44}$ (GeV$^{-4}$)&	$\mathbb{C}_{24}$ (GeV$^{-4}$)&	$\mathbb{C}_{12}$ (GeV$^{-4}$)&	$\mathbb{C}_{14}$ (GeV$^{-4}$)\\
$-0.172\pm0.0216$&	$-1.05\pm0.251$&	$-0.548\pm0.134$&	$-0.734\pm0.188$&	$0.175\pm0.0534$&	$-0.0352\pm0.0163$\\
  \hline
\end{tabular}
}
\end{table*}
Two pronounced peaks around the $\bar{D}^\ast K^\ast$ threshold  are obtained. One can notice that the signal of $P$-wave is much stronger than that of the $S$-wave, this can naturally explain why the angular distribution of $D^-K^+$ in experiments is overwhelmed by the $P$-wave structure~\cite{LHCb:2020bls,LHCb:2020pxc}.

We then study the pole structures of the production  $\mathcal{U}$-matrix in different Riemann sheets (the poles of the $\mathcal{U}$-matrix are the same as the $T$-matrix in this sense),
which can be achieved through the analytical continuation of the Green's functions in Eq.~\eqref{eq:GreenF}:
\begin{eqnarray}
\mathcal{G}_1(E)&\to& \mathcal{G}_1(E)+\zeta_1\frac{i\mu_1p_1}{4\pi^2},\nonumber\\
\mathcal{G}_2(E)&\to& \mathcal{G}_2(E)+\zeta_2\frac{i\mu_2p_2}{4\pi^2}.
\end{eqnarray}
Two channel coupling generally introduces four Riemann sheets~\cite{Frazer:1964zz}, then the four Riemann sheets $(\zeta_1,\zeta_2)$ in the complex energy plane are defined as
\begin{eqnarray}
{\rm Sheet ~I}:(\zeta_1,\zeta_2)&=&(0,0),\nonumber\\
{\rm Sheet ~II}:(\zeta_1,\zeta_2)&=&(1,0),\nonumber\\
{\rm Sheet ~III}:(\zeta_1,\zeta_2)&=&(1,1),\nonumber\\
{\rm Sheet ~IV}:(\zeta_1,\zeta_2)&=&(0,1),
\end{eqnarray}
such as the physical sheet is denoted as $(0,0)$ in this definition. This is analogous to the commonly used notations in complex momentum plane~\cite{Frazer:1964zz}. Considering a lower channel-$1$ with particles $A$ and $B$, and a higher channel-$2$ with particles $C$ and $D$, a pole in Sheet II near $m_C+m_D$ corresponds to a bound state of $CD$ (channel-$2$), such a pole is manifested as a resonance in channel-$1$ ($AB$) scattering process. For more general discussions on the pole distributions and classifications in different Riemann sheets, we refer to~\cite{Frazer:1964zz,Eden:1964zz,Badalian:1981xj}.

The $S$- and $P$-wave peaks correspond to two poles of the $\mathcal{U}$-matrix in the Sheet II. The
masses and widths of the $S$-wave state with $I(J^P)=0(0^+)$ and
$P$-wave state with $0(1^-)$ are extracted from the pole positions
$E=m-i\Gamma/2$. Their values are insensitive to different cutoffs. We obtain
\begin{eqnarray}
[m,\Gamma]_{0^+}&\simeq&[2873.6^{+11.3}_{-13.5},72.1^{+10.5}_{-9.8}] ~\rm{MeV},\nonumber\\
{[m,\Gamma]_{1^-}}&\simeq&[2892.7^{+6.8}_{-10.2},54.2^{+8.2}_{-1.5}] ~\rm{MeV}.
\end{eqnarray}
Therefore, they correspond to the two bound states of $\bar{D}^\ast K^\ast$ in $S$- and $P$-waves, respectively.
We find the masses of the $0^+$ and $1^-$ states are very consistent
with the $X_0(2900)$ and
$X_1(2900)$~\cite{LHCb:2020bls,LHCb:2020pxc}, respectively, while
their widths deviate a lot. The analyses from the experimental work
give $\Gamma_{X_1}>\Gamma_{X_0}$~\cite{LHCb:2020bls,LHCb:2020pxc},
whereas this relation is reversed in our results. The
result in Ref.~\cite{Huang:2020ptc} agrees with ours.

Inspecting the $S$- and $P$-wave peaks in Fig.~\ref{FitPlot}, one
may intuitively infer that the width of $P$-wave state is larger
than that of the $S$-wave. But as we mentioned above, we do have
$\Gamma_{0^+}>\Gamma_{1^-}$ from the pole analyses. This is because
the $P$-wave production vertex is momentum dependent, which broadens
the line shape, but the pole position cannot be changed.

The experimental data may consist of multi substructures, similar to the
stories of the $P_c$ states [the $P_c(4450)$ observed in 2015 by the
LHCb~\cite{LHCb:2015yax} was proved to contain two distinct states
$P_c(4440)$ and $P_c(4457)$ in 2019~\cite{LHCb:2019kea}]. The $S$-
and $P$-wave multiplets for the $\bar{D}^\ast K^\ast$ system are
$[0^+,1^+,2^+]$ and $\{1^-,[0^-,1^-,2^-],[1^-,2^-,3^-]\}$,
respectively. The present experimental data can be well reproduced
with the inclusion of the $0^+$ and $1^-$ states, but the other
states in the $S$- and $P$-wave  multiplets might also exist. In
2010, Molina {\it et al} predicted the existence of the $S$-wave
multiplets in $\bar{D}^\ast K^\ast$ system~\cite{Molina:2010tx}, in
which the $0^+$ state mass is roughly consistent with our result.
The present resolution cannot distinguish the spin-spin and
spin-orbital interactions, thus more experimental data are still
needed. However, one should also note that these possibly existed
substructures might be hard to be discerned in experiments given
that their widthes were similar to those of the $X_{0,1}(2900)$ states.

\section{Summary}\label{Sum}

We have investigated the $D^-K^+$ invariant mass distributions for
the decay process $B^+\to D^+D^-K^+$, where the $X_0(2900)$ and
$X_1(2900)$ were observed by the LHCb
Collaboration~\cite{LHCb:2020bls,LHCb:2020pxc}. We study the $S$- and $P$-wave $\bar{D}^\ast K^\ast$
interactions in a coupled-channel formalism via mimicking
the $\chi$EFT. The short- and
long-distance forces are both incorporated in our calculations.

The event distributions can be well described, and two sharp peaks
that correspond to the $\bar{D}^\ast K^\ast$ molecular states with
$0(0^+)$ and $0(1^-)$  are obtained. Their masses are in good
agreement with the $X_0(2900)$ and $X_1(2900)$, respectively, but
the width of $0^+$ state is larger than that of the $1^-$. We can
simultaneously reproduce the  $X_0(2900)$ and $X_1(2900)$ in an
unified framework. Our calculations
support the molecular interpretations for these two states. In other
words, the $X_1(2900)$ may be the $P$-wave excitation of the ground
state $X_0(2900)$. If this conclusion is verified in future, it
shall be the first time that a ground-state hadronic molecule and
its orbital excitation are synchronously observed in experiments.

However, the present experimental data cannot pin down the fine
structures behind the $\bar{D}^\ast K^\ast$ interactions, i.e., we
still cannot ascertain whether the other members in the $S$- and
$P$-wave multiplets exist. The single whole peak now might contain
multi substructures. In order to resolve the structures hidden in the
peak, some refined measurements in this decay channel are still
needed.

\section*{Acknowledgments}
B. Wang is very grateful to L. Meng for helpful discussions and
carefully reading the manuscript.  This work is supported
by the National Natural Science Foundation
of China under Grants No. 12105072, No. 11975033 and No. 12070131001.
B.W is also supported by the Youth Funds of Hebei Province (No. A2021201027) and the
Start-up Funds for Young Talents of Hebei University (No. 521100221021).

\bibliography{ref}

\begin{thebibliography}{77}%
\makeatletter
\providecommand \@ifxundefined [1]{%
 \@ifx{#1\undefined}
}%
\providecommand \@ifnum [1]{%
 \ifnum #1\expandafter \@firstoftwo
 \else \expandafter \@secondoftwo
 \fi
}%
\providecommand \@ifx [1]{%
 \ifx #1\expandafter \@firstoftwo
 \else \expandafter \@secondoftwo
 \fi
}%
\providecommand \natexlab [1]{#1}%
\providecommand \enquote  [1]{``#1''}%
\providecommand \bibnamefont  [1]{#1}%
\providecommand \bibfnamefont [1]{#1}%
\providecommand \citenamefont [1]{#1}%
\providecommand \href@noop [0]{\@secondoftwo}%
\providecommand \href [0]{\begingroup \@sanitize@url \@href}%
\providecommand \@href[1]{\@@startlink{#1}\@@href}%
\providecommand \@@href[1]{\endgroup#1\@@endlink}%
\providecommand \@sanitize@url [0]{\catcode `\\12\catcode `\$12\catcode
  `\&12\catcode `\#12\catcode `\^12\catcode `\_12\catcode `\%12\relax}%
\providecommand \@@startlink[1]{}%
\providecommand \@@endlink[0]{}%
\providecommand \url  [0]{\begingroup\@sanitize@url \@url }%
\providecommand \@url [1]{\endgroup\@href {#1}{\urlprefix }}%
\providecommand \urlprefix  [0]{URL }%
\providecommand \Eprint [0]{\href }%
\providecommand \doibase [0]{http://dx.doi.org/}%
\providecommand \selectlanguage [0]{\@gobble}%
\providecommand \bibinfo  [0]{\@secondoftwo}%
\providecommand \bibfield  [0]{\@secondoftwo}%
\providecommand \translation [1]{[#1]}%
\providecommand \BibitemOpen [0]{}%
\providecommand \bibitemStop [0]{}%
\providecommand \bibitemNoStop [0]{.\EOS\space}%
\providecommand \EOS [0]{\spacefactor3000\relax}%
\providecommand \BibitemShut  [1]{\csname bibitem#1\endcsname}%
\let\auto@bib@innerbib\@empty
\bibitem [{\citenamefont {Aaij}\ \emph
  {et~al.}(2020{\natexlab{a}})\citenamefont {Aaij} \emph
  {et~al.}}]{LHCb:2020bls}%
  \BibitemOpen
  \bibfield  {author} {\bibinfo {author} {\bibfnamefont {R.}~\bibnamefont
  {Aaij}} \emph {et~al.} (\bibinfo {collaboration} {LHCb}),\ }\href {\doibase
  10.1103/PhysRevLett.125.242001} {\bibfield  {journal} {\bibinfo  {journal}
  {Phys. Rev. Lett.}\ }\textbf {\bibinfo {volume} {125}},\ \bibinfo {pages}
  {242001} (\bibinfo {year} {2020}{\natexlab{a}})},\ \Eprint
  {http://arxiv.org/abs/2009.00025} {arXiv:2009.00025 [hep-ex]} \BibitemShut
  {NoStop}%
\bibitem [{\citenamefont {Aaij}\ \emph
  {et~al.}(2020{\natexlab{b}})\citenamefont {Aaij} \emph
  {et~al.}}]{LHCb:2020pxc}%
  \BibitemOpen
  \bibfield  {author} {\bibinfo {author} {\bibfnamefont {R.}~\bibnamefont
  {Aaij}} \emph {et~al.} (\bibinfo {collaboration} {LHCb}),\ }\href {\doibase
  10.1103/PhysRevD.102.112003} {\bibfield  {journal} {\bibinfo  {journal}
  {Phys. Rev. D}\ }\textbf {\bibinfo {volume} {102}},\ \bibinfo {pages}
  {112003} (\bibinfo {year} {2020}{\natexlab{b}})},\ \Eprint
  {http://arxiv.org/abs/2009.00026} {arXiv:2009.00026 [hep-ex]} \BibitemShut
  {NoStop}%
\bibitem [{\citenamefont {Chen}\ \emph {et~al.}(2020)\citenamefont {Chen},
  \citenamefont {Chen}, \citenamefont {Dong},\ and\ \citenamefont
  {Su}}]{Chen:2020aos}%
  \BibitemOpen
  \bibfield  {author} {\bibinfo {author} {\bibfnamefont {H.-X.}\ \bibnamefont
  {Chen}}, \bibinfo {author} {\bibfnamefont {W.}~\bibnamefont {Chen}}, \bibinfo
  {author} {\bibfnamefont {R.-R.}\ \bibnamefont {Dong}}, \ and\ \bibinfo
  {author} {\bibfnamefont {N.}~\bibnamefont {Su}},\ }\href {\doibase
  10.1088/0256-307X/37/10/101201} {\bibfield  {journal} {\bibinfo  {journal}
  {Chin. Phys. Lett.}\ }\textbf {\bibinfo {volume} {37}},\ \bibinfo {pages}
  {101201} (\bibinfo {year} {2020})},\ \Eprint
  {http://arxiv.org/abs/2008.07516} {arXiv:2008.07516 [hep-ph]} \BibitemShut
  {NoStop}%
\bibitem [{\citenamefont {He}\ and\ \citenamefont {Chen}(2021)}]{He:2020btl}%
  \BibitemOpen
  \bibfield  {author} {\bibinfo {author} {\bibfnamefont {J.}~\bibnamefont
  {He}}\ and\ \bibinfo {author} {\bibfnamefont {D.-Y.}\ \bibnamefont {Chen}},\
  }\href {\doibase 10.1088/1674-1137/abeda8} {\bibfield  {journal} {\bibinfo
  {journal} {Chin. Phys. C}\ }\textbf {\bibinfo {volume} {45}},\ \bibinfo
  {pages} {063102} (\bibinfo {year} {2021})},\ \Eprint
  {http://arxiv.org/abs/2008.07782} {arXiv:2008.07782 [hep-ph]} \BibitemShut
  {NoStop}%
\bibitem [{\citenamefont {Liu}\ \emph {et~al.}(2020{\natexlab{a}})\citenamefont
  {Liu}, \citenamefont {Xie},\ and\ \citenamefont {Geng}}]{Liu:2020nil}%
  \BibitemOpen
  \bibfield  {author} {\bibinfo {author} {\bibfnamefont {M.-Z.}\ \bibnamefont
  {Liu}}, \bibinfo {author} {\bibfnamefont {J.-J.}\ \bibnamefont {Xie}}, \ and\
  \bibinfo {author} {\bibfnamefont {L.-S.}\ \bibnamefont {Geng}},\ }\href
  {\doibase 10.1103/PhysRevD.102.091502} {\bibfield  {journal} {\bibinfo
  {journal} {Phys. Rev. D}\ }\textbf {\bibinfo {volume} {102}},\ \bibinfo
  {pages} {091502} (\bibinfo {year} {2020}{\natexlab{a}})},\ \Eprint
  {http://arxiv.org/abs/2008.07389} {arXiv:2008.07389 [hep-ph]} \BibitemShut
  {NoStop}%
\bibitem [{\citenamefont {Hu}\ \emph {et~al.}(2021)\citenamefont {Hu},
  \citenamefont {Lao}, \citenamefont {Ling},\ and\ \citenamefont
  {Wang}}]{Hu:2020mxp}%
  \BibitemOpen
  \bibfield  {author} {\bibinfo {author} {\bibfnamefont {M.-W.}\ \bibnamefont
  {Hu}}, \bibinfo {author} {\bibfnamefont {X.-Y.}\ \bibnamefont {Lao}},
  \bibinfo {author} {\bibfnamefont {P.}~\bibnamefont {Ling}}, \ and\ \bibinfo
  {author} {\bibfnamefont {Q.}~\bibnamefont {Wang}},\ }\href {\doibase
  10.1088/1674-1137/abcfaa} {\bibfield  {journal} {\bibinfo  {journal} {Chin.
  Phys. C}\ }\textbf {\bibinfo {volume} {45}},\ \bibinfo {pages} {021003}
  (\bibinfo {year} {2021})},\ \Eprint {http://arxiv.org/abs/2008.06894}
  {arXiv:2008.06894 [hep-ph]} \BibitemShut {NoStop}%
\bibitem [{\citenamefont {Agaev}\ \emph {et~al.}(2020)\citenamefont {Agaev},
  \citenamefont {Azizi},\ and\ \citenamefont {Sundu}}]{Agaev:2020nrc}%
  \BibitemOpen
  \bibfield  {author} {\bibinfo {author} {\bibfnamefont {S.~S.}\ \bibnamefont
  {Agaev}}, \bibinfo {author} {\bibfnamefont {K.}~\bibnamefont {Azizi}}, \ and\
  \bibinfo {author} {\bibfnamefont {H.}~\bibnamefont {Sundu}},\ }\href@noop {}
  {\  (\bibinfo {year} {2020})},\ \Eprint {http://arxiv.org/abs/2008.13027}
  {arXiv:2008.13027 [hep-ph]} \BibitemShut {NoStop}%
\bibitem [{\citenamefont {Karliner}\ and\ \citenamefont
  {Rosner}(2020)}]{Karliner:2020vsi}%
  \BibitemOpen
  \bibfield  {author} {\bibinfo {author} {\bibfnamefont {M.}~\bibnamefont
  {Karliner}}\ and\ \bibinfo {author} {\bibfnamefont {J.~L.}\ \bibnamefont
  {Rosner}},\ }\href {\doibase 10.1103/PhysRevD.102.094016} {\bibfield
  {journal} {\bibinfo  {journal} {Phys. Rev. D}\ }\textbf {\bibinfo {volume}
  {102}},\ \bibinfo {pages} {094016} (\bibinfo {year} {2020})},\ \Eprint
  {http://arxiv.org/abs/2008.05993} {arXiv:2008.05993 [hep-ph]} \BibitemShut
  {NoStop}%
\bibitem [{\citenamefont {He}\ \emph {et~al.}(2020)\citenamefont {He},
  \citenamefont {Wang},\ and\ \citenamefont {Zhu}}]{He:2020jna}%
  \BibitemOpen
  \bibfield  {author} {\bibinfo {author} {\bibfnamefont {X.-G.}\ \bibnamefont
  {He}}, \bibinfo {author} {\bibfnamefont {W.}~\bibnamefont {Wang}}, \ and\
  \bibinfo {author} {\bibfnamefont {R.}~\bibnamefont {Zhu}},\ }\href {\doibase
  10.1140/epjc/s10052-020-08597-1} {\bibfield  {journal} {\bibinfo  {journal}
  {Eur. Phys. J. C}\ }\textbf {\bibinfo {volume} {80}},\ \bibinfo {pages}
  {1026} (\bibinfo {year} {2020})},\ \Eprint {http://arxiv.org/abs/2008.07145}
  {arXiv:2008.07145 [hep-ph]} \BibitemShut {NoStop}%
\bibitem [{\citenamefont {Wang}(2020)}]{Wang:2020xyc}%
  \BibitemOpen
  \bibfield  {author} {\bibinfo {author} {\bibfnamefont {Z.-G.}\ \bibnamefont
  {Wang}},\ }\href {\doibase 10.1142/S0217751X20501870} {\bibfield  {journal}
  {\bibinfo  {journal} {Int. J. Mod. Phys. A}\ }\textbf {\bibinfo {volume}
  {35}},\ \bibinfo {pages} {2050187} (\bibinfo {year} {2020})},\ \Eprint
  {http://arxiv.org/abs/2008.07833} {arXiv:2008.07833 [hep-ph]} \BibitemShut
  {NoStop}%
\bibitem [{\citenamefont {Zhang}(2021)}]{Zhang:2020oze}%
  \BibitemOpen
  \bibfield  {author} {\bibinfo {author} {\bibfnamefont {J.-R.}\ \bibnamefont
  {Zhang}},\ }\href {\doibase 10.1103/PhysRevD.103.054019} {\bibfield
  {journal} {\bibinfo  {journal} {Phys. Rev. D}\ }\textbf {\bibinfo {volume}
  {103}},\ \bibinfo {pages} {054019} (\bibinfo {year} {2021})},\ \Eprint
  {http://arxiv.org/abs/2008.07295} {arXiv:2008.07295 [hep-ph]} \BibitemShut
  {NoStop}%
\bibitem [{\citenamefont {Wang}\ \emph {et~al.}(2021)\citenamefont {Wang},
  \citenamefont {Meng}, \citenamefont {Xiao}, \citenamefont {Oka},\ and\
  \citenamefont {Zhu}}]{Wang:2020prk}%
  \BibitemOpen
  \bibfield  {author} {\bibinfo {author} {\bibfnamefont {G.-J.}\ \bibnamefont
  {Wang}}, \bibinfo {author} {\bibfnamefont {L.}~\bibnamefont {Meng}}, \bibinfo
  {author} {\bibfnamefont {L.-Y.}\ \bibnamefont {Xiao}}, \bibinfo {author}
  {\bibfnamefont {M.}~\bibnamefont {Oka}}, \ and\ \bibinfo {author}
  {\bibfnamefont {S.-L.}\ \bibnamefont {Zhu}},\ }\href {\doibase
  10.1140/epjc/s10052-021-08978-0} {\bibfield  {journal} {\bibinfo  {journal}
  {Eur. Phys. J. C}\ }\textbf {\bibinfo {volume} {81}},\ \bibinfo {pages} {188}
  (\bibinfo {year} {2021})},\ \Eprint {http://arxiv.org/abs/2010.09395}
  {arXiv:2010.09395 [hep-ph]} \BibitemShut {NoStop}%
\bibitem [{\citenamefont {Liu}\ \emph {et~al.}(2020{\natexlab{b}})\citenamefont
  {Liu}, \citenamefont {Yan}, \citenamefont {Ke}, \citenamefont {Li},\ and\
  \citenamefont {Xie}}]{Liu:2020orv}%
  \BibitemOpen
  \bibfield  {author} {\bibinfo {author} {\bibfnamefont {X.-H.}\ \bibnamefont
  {Liu}}, \bibinfo {author} {\bibfnamefont {M.-J.}\ \bibnamefont {Yan}},
  \bibinfo {author} {\bibfnamefont {H.-W.}\ \bibnamefont {Ke}}, \bibinfo
  {author} {\bibfnamefont {G.}~\bibnamefont {Li}}, \ and\ \bibinfo {author}
  {\bibfnamefont {J.-J.}\ \bibnamefont {Xie}},\ }\href {\doibase
  10.1140/epjc/s10052-020-08762-6} {\bibfield  {journal} {\bibinfo  {journal}
  {Eur. Phys. J. C}\ }\textbf {\bibinfo {volume} {80}},\ \bibinfo {pages}
  {1178} (\bibinfo {year} {2020}{\natexlab{b}})},\ \Eprint
  {http://arxiv.org/abs/2008.07190} {arXiv:2008.07190 [hep-ph]} \BibitemShut
  {NoStop}%
\bibitem [{\citenamefont {Burns}\ and\ \citenamefont
  {Swanson}(2021{\natexlab{a}})}]{Burns:2020epm}%
  \BibitemOpen
  \bibfield  {author} {\bibinfo {author} {\bibfnamefont {T.~J.}\ \bibnamefont
  {Burns}}\ and\ \bibinfo {author} {\bibfnamefont {E.~S.}\ \bibnamefont
  {Swanson}},\ }\href {\doibase 10.1016/j.physletb.2020.136057} {\bibfield
  {journal} {\bibinfo  {journal} {Phys. Lett. B}\ }\textbf {\bibinfo {volume}
  {813}},\ \bibinfo {pages} {136057} (\bibinfo {year} {2021}{\natexlab{a}})},\
  \Eprint {http://arxiv.org/abs/2008.12838} {arXiv:2008.12838 [hep-ph]}
  \BibitemShut {NoStop}%
\bibitem [{\citenamefont {Huang}\ \emph {et~al.}(2020)\citenamefont {Huang},
  \citenamefont {Lu}, \citenamefont {Xie},\ and\ \citenamefont
  {Geng}}]{Huang:2020ptc}%
  \BibitemOpen
  \bibfield  {author} {\bibinfo {author} {\bibfnamefont {Y.}~\bibnamefont
  {Huang}}, \bibinfo {author} {\bibfnamefont {J.-X.}\ \bibnamefont {Lu}},
  \bibinfo {author} {\bibfnamefont {J.-J.}\ \bibnamefont {Xie}}, \ and\
  \bibinfo {author} {\bibfnamefont {L.-S.}\ \bibnamefont {Geng}},\ }\href
  {\doibase 10.1140/epjc/s10052-020-08516-4} {\bibfield  {journal} {\bibinfo
  {journal} {Eur. Phys. J. C}\ }\textbf {\bibinfo {volume} {80}},\ \bibinfo
  {pages} {973} (\bibinfo {year} {2020})},\ \Eprint
  {http://arxiv.org/abs/2008.07959} {arXiv:2008.07959 [hep-ph]} \BibitemShut
  {NoStop}%
\bibitem [{\citenamefont {Chen}\ \emph
  {et~al.}(2021{\natexlab{a}})\citenamefont {Chen}, \citenamefont {Han},
  \citenamefont {L\"u}, \citenamefont {Wang},\ and\ \citenamefont
  {Yu}}]{Chen:2020eyu}%
  \BibitemOpen
  \bibfield  {author} {\bibinfo {author} {\bibfnamefont {Y.-K.}\ \bibnamefont
  {Chen}}, \bibinfo {author} {\bibfnamefont {J.-J.}\ \bibnamefont {Han}},
  \bibinfo {author} {\bibfnamefont {Q.-F.}\ \bibnamefont {L\"u}}, \bibinfo
  {author} {\bibfnamefont {J.-P.}\ \bibnamefont {Wang}}, \ and\ \bibinfo
  {author} {\bibfnamefont {F.-S.}\ \bibnamefont {Yu}},\ }\href {\doibase
  10.1140/epjc/s10052-021-08857-8} {\bibfield  {journal} {\bibinfo  {journal}
  {Eur. Phys. J. C}\ }\textbf {\bibinfo {volume} {81}},\ \bibinfo {pages} {71}
  (\bibinfo {year} {2021}{\natexlab{a}})},\ \Eprint
  {http://arxiv.org/abs/2009.01182} {arXiv:2009.01182 [hep-ph]} \BibitemShut
  {NoStop}%
\bibitem [{\citenamefont {Burns}\ and\ \citenamefont
  {Swanson}(2021{\natexlab{b}})}]{Burns:2020xne}%
  \BibitemOpen
  \bibfield  {author} {\bibinfo {author} {\bibfnamefont {T.~J.}\ \bibnamefont
  {Burns}}\ and\ \bibinfo {author} {\bibfnamefont {E.~S.}\ \bibnamefont
  {Swanson}},\ }\href {\doibase 10.1103/PhysRevD.103.014004} {\bibfield
  {journal} {\bibinfo  {journal} {Phys. Rev. D}\ }\textbf {\bibinfo {volume}
  {103}},\ \bibinfo {pages} {014004} (\bibinfo {year} {2021}{\natexlab{b}})},\
  \Eprint {http://arxiv.org/abs/2009.05352} {arXiv:2009.05352 [hep-ph]}
  \BibitemShut {NoStop}%
\bibitem [{\citenamefont {Xiao}\ \emph {et~al.}(2021)\citenamefont {Xiao},
  \citenamefont {Chen}, \citenamefont {Dong},\ and\ \citenamefont
  {Meng}}]{Xiao:2020ltm}%
  \BibitemOpen
  \bibfield  {author} {\bibinfo {author} {\bibfnamefont {C.-J.}\ \bibnamefont
  {Xiao}}, \bibinfo {author} {\bibfnamefont {D.-Y.}\ \bibnamefont {Chen}},
  \bibinfo {author} {\bibfnamefont {Y.-B.}\ \bibnamefont {Dong}}, \ and\
  \bibinfo {author} {\bibfnamefont {G.-W.}\ \bibnamefont {Meng}},\ }\href
  {\doibase 10.1103/PhysRevD.103.034004} {\bibfield  {journal} {\bibinfo
  {journal} {Phys. Rev. D}\ }\textbf {\bibinfo {volume} {103}},\ \bibinfo
  {pages} {034004} (\bibinfo {year} {2021})},\ \Eprint
  {http://arxiv.org/abs/2009.14538} {arXiv:2009.14538 [hep-ph]} \BibitemShut
  {NoStop}%
\bibitem [{\citenamefont {Albuquerque}\ \emph {et~al.}(2021)\citenamefont
  {Albuquerque}, \citenamefont {Narison}, \citenamefont {Rabetiarivony},\ and\
  \citenamefont {Randriamanatrika}}]{Albuquerque:2020ugi}%
  \BibitemOpen
  \bibfield  {author} {\bibinfo {author} {\bibfnamefont {R.~M.}\ \bibnamefont
  {Albuquerque}}, \bibinfo {author} {\bibfnamefont {S.}~\bibnamefont
  {Narison}}, \bibinfo {author} {\bibfnamefont {D.}~\bibnamefont
  {Rabetiarivony}}, \ and\ \bibinfo {author} {\bibfnamefont {G.}~\bibnamefont
  {Randriamanatrika}},\ }\href {\doibase 10.1016/j.nuclphysa.2020.122113}
  {\bibfield  {journal} {\bibinfo  {journal} {Nucl. Phys. A}\ }\textbf
  {\bibinfo {volume} {1007}},\ \bibinfo {pages} {122113} (\bibinfo {year}
  {2021})},\ \Eprint {http://arxiv.org/abs/2008.13463} {arXiv:2008.13463
  [hep-ph]} \BibitemShut {NoStop}%
\bibitem [{\citenamefont {L\"u}\ \emph {et~al.}(2020)\citenamefont {L\"u},
  \citenamefont {Chen},\ and\ \citenamefont {Dong}}]{Lu:2020qmp}%
  \BibitemOpen
  \bibfield  {author} {\bibinfo {author} {\bibfnamefont {Q.-F.}\ \bibnamefont
  {L\"u}}, \bibinfo {author} {\bibfnamefont {D.-Y.}\ \bibnamefont {Chen}}, \
  and\ \bibinfo {author} {\bibfnamefont {Y.-B.}\ \bibnamefont {Dong}},\ }\href
  {\doibase 10.1103/PhysRevD.102.074021} {\bibfield  {journal} {\bibinfo
  {journal} {Phys. Rev. D}\ }\textbf {\bibinfo {volume} {102}},\ \bibinfo
  {pages} {074021} (\bibinfo {year} {2020})},\ \Eprint
  {http://arxiv.org/abs/2008.07340} {arXiv:2008.07340 [hep-ph]} \BibitemShut
  {NoStop}%
\bibitem [{\citenamefont {Mutuk}(2021)}]{Mutuk:2020igv}%
  \BibitemOpen
  \bibfield  {author} {\bibinfo {author} {\bibfnamefont {H.}~\bibnamefont
  {Mutuk}},\ }\href {\doibase 10.1088/1361-6471/abeb7f} {\bibfield  {journal}
  {\bibinfo  {journal} {J. Phys. G}\ }\textbf {\bibinfo {volume} {48}},\
  \bibinfo {pages} {055007} (\bibinfo {year} {2021})},\ \Eprint
  {http://arxiv.org/abs/2009.02492} {arXiv:2009.02492 [hep-ph]} \BibitemShut
  {NoStop}%
\bibitem [{\citenamefont {Tan}\ and\ \citenamefont {Ping}(2020)}]{Tan:2020cpu}%
  \BibitemOpen
  \bibfield  {author} {\bibinfo {author} {\bibfnamefont {Y.}~\bibnamefont
  {Tan}}\ and\ \bibinfo {author} {\bibfnamefont {J.}~\bibnamefont {Ping}},\
  }\href@noop {} {\  (\bibinfo {year} {2020})},\ \Eprint
  {http://arxiv.org/abs/2010.04045} {arXiv:2010.04045 [hep-ph]} \BibitemShut
  {NoStop}%
\bibitem [{\citenamefont {Abreu}(2021)}]{Abreu:2020ony}%
  \BibitemOpen
  \bibfield  {author} {\bibinfo {author} {\bibfnamefont {L.~M.}\ \bibnamefont
  {Abreu}},\ }\href {\doibase 10.1103/PhysRevD.103.036013} {\bibfield
  {journal} {\bibinfo  {journal} {Phys. Rev. D}\ }\textbf {\bibinfo {volume}
  {103}},\ \bibinfo {pages} {036013} (\bibinfo {year} {2021})},\ \Eprint
  {http://arxiv.org/abs/2010.14955} {arXiv:2010.14955 [hep-ph]} \BibitemShut
  {NoStop}%
\bibitem [{\citenamefont {Qi}\ \emph {et~al.}(2021)\citenamefont {Qi},
  \citenamefont {Wang}, \citenamefont {Zhang},\ and\ \citenamefont
  {Guo}}]{Qi:2021iyv}%
  \BibitemOpen
  \bibfield  {author} {\bibinfo {author} {\bibfnamefont {J.-J.}\ \bibnamefont
  {Qi}}, \bibinfo {author} {\bibfnamefont {Z.-Y.}\ \bibnamefont {Wang}},
  \bibinfo {author} {\bibfnamefont {Z.-F.}\ \bibnamefont {Zhang}}, \ and\
  \bibinfo {author} {\bibfnamefont {X.-H.}\ \bibnamefont {Guo}},\ }\href@noop
  {} {\  (\bibinfo {year} {2021})},\ \Eprint {http://arxiv.org/abs/2101.06688}
  {arXiv:2101.06688 [hep-ph]} \BibitemShut {NoStop}%
\bibitem [{\citenamefont {Chen}(2021)}]{Chen:2021erj}%
  \BibitemOpen
  \bibfield  {author} {\bibinfo {author} {\bibfnamefont {H.-X.}\ \bibnamefont
  {Chen}},\ }\href@noop {} {\  (\bibinfo {year} {2021})},\ \Eprint
  {http://arxiv.org/abs/2103.08586} {arXiv:2103.08586 [hep-ph]} \BibitemShut
  {NoStop}%
\bibitem [{\citenamefont {Hsiao}\ and\ \citenamefont
  {Yu}(2021)}]{Hsiao:2021tyq}%
  \BibitemOpen
  \bibfield  {author} {\bibinfo {author} {\bibfnamefont {Y.-K.}\ \bibnamefont
  {Hsiao}}\ and\ \bibinfo {author} {\bibfnamefont {Y.}~\bibnamefont {Yu}},\
  }\href@noop {} {\  (\bibinfo {year} {2021})},\ \Eprint
  {http://arxiv.org/abs/2104.01296} {arXiv:2104.01296 [hep-ph]} \BibitemShut
  {NoStop}%
\bibitem [{\citenamefont {Duan}\ \emph {et~al.}(2021)\citenamefont {Duan},
  \citenamefont {Wang}, \citenamefont {Li},\ and\ \citenamefont
  {Liu}}]{Duan:2021bna}%
  \BibitemOpen
  \bibfield  {author} {\bibinfo {author} {\bibfnamefont {M.-X.}\ \bibnamefont
  {Duan}}, \bibinfo {author} {\bibfnamefont {J.-Z.}\ \bibnamefont {Wang}},
  \bibinfo {author} {\bibfnamefont {Y.-S.}\ \bibnamefont {Li}}, \ and\ \bibinfo
  {author} {\bibfnamefont {X.}~\bibnamefont {Liu}},\ }\href@noop {} {\
  (\bibinfo {year} {2021})},\ \Eprint {http://arxiv.org/abs/2104.09132}
  {arXiv:2104.09132 [hep-ph]} \BibitemShut {NoStop}%
\bibitem [{\citenamefont {Kong}\ \emph {et~al.}(2021)\citenamefont {Kong},
  \citenamefont {Zhu}, \citenamefont {Song},\ and\ \citenamefont
  {He}}]{Kong:2021ohg}%
  \BibitemOpen
  \bibfield  {author} {\bibinfo {author} {\bibfnamefont {S.-Y.}\ \bibnamefont
  {Kong}}, \bibinfo {author} {\bibfnamefont {J.-T.}\ \bibnamefont {Zhu}},
  \bibinfo {author} {\bibfnamefont {D.}~\bibnamefont {Song}}, \ and\ \bibinfo
  {author} {\bibfnamefont {J.}~\bibnamefont {He}},\ }\href@noop {} {\
  (\bibinfo {year} {2021})},\ \Eprint {http://arxiv.org/abs/2106.07272}
  {arXiv:2106.07272 [hep-ph]} \BibitemShut {NoStop}%
\bibitem [{\citenamefont {Dong}\ and\ \citenamefont
  {Zou}(2021)}]{Dong:2020rgs}%
  \BibitemOpen
  \bibfield  {author} {\bibinfo {author} {\bibfnamefont {X.-K.}\ \bibnamefont
  {Dong}}\ and\ \bibinfo {author} {\bibfnamefont {B.-S.}\ \bibnamefont {Zou}},\
  }\href {\doibase 10.1140/epja/s10050-021-00442-7} {\bibfield  {journal}
  {\bibinfo  {journal} {Eur. Phys. J. A}\ }\textbf {\bibinfo {volume} {57}},\
  \bibinfo {pages} {139} (\bibinfo {year} {2021})},\ \Eprint
  {http://arxiv.org/abs/2009.11619} {arXiv:2009.11619 [hep-ph]} \BibitemShut
  {NoStop}%
\bibitem [{\citenamefont {Bondar}\ and\ \citenamefont
  {Milstein}(2020)}]{Bondar:2020eoa}%
  \BibitemOpen
  \bibfield  {author} {\bibinfo {author} {\bibfnamefont {A.~E.}\ \bibnamefont
  {Bondar}}\ and\ \bibinfo {author} {\bibfnamefont {A.~I.}\ \bibnamefont
  {Milstein}},\ }\href {\doibase 10.1007/JHEP12(2020)015} {\bibfield  {journal}
  {\bibinfo  {journal} {JHEP}\ }\textbf {\bibinfo {volume} {12}},\ \bibinfo
  {pages} {015} (\bibinfo {year} {2020})},\ \Eprint
  {http://arxiv.org/abs/2008.13337} {arXiv:2008.13337 [hep-ph]} \BibitemShut
  {NoStop}%
\bibitem [{\citenamefont {Chen}\ \emph
  {et~al.}(2021{\natexlab{b}})\citenamefont {Chen}, \citenamefont {Qi},\ and\
  \citenamefont {Zheng}}]{Chen:2021tad}%
  \BibitemOpen
  \bibfield  {author} {\bibinfo {author} {\bibfnamefont {H.}~\bibnamefont
  {Chen}}, \bibinfo {author} {\bibfnamefont {H.-R.}\ \bibnamefont {Qi}}, \ and\
  \bibinfo {author} {\bibfnamefont {H.-Q.}\ \bibnamefont {Zheng}},\ }\href
  {\doibase 10.1140/epjc/s10052-021-09603-w} {\bibfield  {journal} {\bibinfo
  {journal} {Eur. Phys. J. C}\ }\textbf {\bibinfo {volume} {81}},\ \bibinfo
  {pages} {812} (\bibinfo {year} {2021}{\natexlab{b}})},\ \Eprint
  {http://arxiv.org/abs/2108.02387} {arXiv:2108.02387 [hep-ph]} \BibitemShut
  {NoStop}%
\bibitem [{\citenamefont {Chen}\ \emph {et~al.}(2016)\citenamefont {Chen},
  \citenamefont {Chen}, \citenamefont {Liu},\ and\ \citenamefont
  {Zhu}}]{Chen:2016qju}%
  \BibitemOpen
  \bibfield  {author} {\bibinfo {author} {\bibfnamefont {H.-X.}\ \bibnamefont
  {Chen}}, \bibinfo {author} {\bibfnamefont {W.}~\bibnamefont {Chen}}, \bibinfo
  {author} {\bibfnamefont {X.}~\bibnamefont {Liu}}, \ and\ \bibinfo {author}
  {\bibfnamefont {S.-L.}\ \bibnamefont {Zhu}},\ }\href {\doibase
  10.1016/j.physrep.2016.05.004} {\bibfield  {journal} {\bibinfo  {journal}
  {Phys. Rept.}\ }\textbf {\bibinfo {volume} {639}},\ \bibinfo {pages} {1}
  (\bibinfo {year} {2016})},\ \Eprint {http://arxiv.org/abs/1601.02092}
  {arXiv:1601.02092 [hep-ph]} \BibitemShut {NoStop}%
\bibitem [{\citenamefont {Guo}\ \emph {et~al.}(2018)\citenamefont {Guo},
  \citenamefont {Hanhart}, \citenamefont {Mei\ss{}ner}, \citenamefont {Wang},
  \citenamefont {Zhao},\ and\ \citenamefont {Zou}}]{Guo:2017jvc}%
  \BibitemOpen
  \bibfield  {author} {\bibinfo {author} {\bibfnamefont {F.-K.}\ \bibnamefont
  {Guo}}, \bibinfo {author} {\bibfnamefont {C.}~\bibnamefont {Hanhart}},
  \bibinfo {author} {\bibfnamefont {U.-G.}\ \bibnamefont {Mei\ss{}ner}},
  \bibinfo {author} {\bibfnamefont {Q.}~\bibnamefont {Wang}}, \bibinfo {author}
  {\bibfnamefont {Q.}~\bibnamefont {Zhao}}, \ and\ \bibinfo {author}
  {\bibfnamefont {B.-S.}\ \bibnamefont {Zou}},\ }\href {\doibase
  10.1103/RevModPhys.90.015004} {\bibfield  {journal} {\bibinfo  {journal}
  {Rev. Mod. Phys.}\ }\textbf {\bibinfo {volume} {90}},\ \bibinfo {pages}
  {015004} (\bibinfo {year} {2018})},\ \Eprint
  {http://arxiv.org/abs/1705.00141} {arXiv:1705.00141 [hep-ph]} \BibitemShut
  {NoStop}%
\bibitem [{\citenamefont {Liu}\ \emph {et~al.}(2019)\citenamefont {Liu},
  \citenamefont {Chen}, \citenamefont {Chen}, \citenamefont {Liu},\ and\
  \citenamefont {Zhu}}]{Liu:2019zoy}%
  \BibitemOpen
  \bibfield  {author} {\bibinfo {author} {\bibfnamefont {Y.-R.}\ \bibnamefont
  {Liu}}, \bibinfo {author} {\bibfnamefont {H.-X.}\ \bibnamefont {Chen}},
  \bibinfo {author} {\bibfnamefont {W.}~\bibnamefont {Chen}}, \bibinfo {author}
  {\bibfnamefont {X.}~\bibnamefont {Liu}}, \ and\ \bibinfo {author}
  {\bibfnamefont {S.-L.}\ \bibnamefont {Zhu}},\ }\href {\doibase
  10.1016/j.ppnp.2019.04.003} {\bibfield  {journal} {\bibinfo  {journal} {Prog.
  Part. Nucl. Phys.}\ }\textbf {\bibinfo {volume} {107}},\ \bibinfo {pages}
  {237} (\bibinfo {year} {2019})},\ \Eprint {http://arxiv.org/abs/1903.11976}
  {arXiv:1903.11976 [hep-ph]} \BibitemShut {NoStop}%
\bibitem [{\citenamefont {Lebed}\ \emph {et~al.}(2017)\citenamefont {Lebed},
  \citenamefont {Mitchell},\ and\ \citenamefont {Swanson}}]{Lebed:2016hpi}%
  \BibitemOpen
  \bibfield  {author} {\bibinfo {author} {\bibfnamefont {R.~F.}\ \bibnamefont
  {Lebed}}, \bibinfo {author} {\bibfnamefont {R.~E.}\ \bibnamefont {Mitchell}},
  \ and\ \bibinfo {author} {\bibfnamefont {E.~S.}\ \bibnamefont {Swanson}},\
  }\href {\doibase 10.1016/j.ppnp.2016.11.003} {\bibfield  {journal} {\bibinfo
  {journal} {Prog. Part. Nucl. Phys.}\ }\textbf {\bibinfo {volume} {93}},\
  \bibinfo {pages} {143} (\bibinfo {year} {2017})},\ \Eprint
  {http://arxiv.org/abs/1610.04528} {arXiv:1610.04528 [hep-ph]} \BibitemShut
  {NoStop}%
\bibitem [{\citenamefont {Esposito}\ \emph {et~al.}(2017)\citenamefont
  {Esposito}, \citenamefont {Pilloni},\ and\ \citenamefont
  {Polosa}}]{Esposito:2016noz}%
  \BibitemOpen
  \bibfield  {author} {\bibinfo {author} {\bibfnamefont {A.}~\bibnamefont
  {Esposito}}, \bibinfo {author} {\bibfnamefont {A.}~\bibnamefont {Pilloni}}, \
  and\ \bibinfo {author} {\bibfnamefont {A.~D.}\ \bibnamefont {Polosa}},\
  }\href {\doibase 10.1016/j.physrep.2016.11.002} {\bibfield  {journal}
  {\bibinfo  {journal} {Phys. Rept.}\ }\textbf {\bibinfo {volume} {668}},\
  \bibinfo {pages} {1} (\bibinfo {year} {2017})},\ \Eprint
  {http://arxiv.org/abs/1611.07920} {arXiv:1611.07920 [hep-ph]} \BibitemShut
  {NoStop}%
\bibitem [{\citenamefont {Brambilla}\ \emph {et~al.}(2020)\citenamefont
  {Brambilla}, \citenamefont {Eidelman}, \citenamefont {Hanhart}, \citenamefont
  {Nefediev}, \citenamefont {Shen}, \citenamefont {Thomas}, \citenamefont
  {Vairo},\ and\ \citenamefont {Yuan}}]{Brambilla:2019esw}%
  \BibitemOpen
  \bibfield  {author} {\bibinfo {author} {\bibfnamefont {N.}~\bibnamefont
  {Brambilla}}, \bibinfo {author} {\bibfnamefont {S.}~\bibnamefont {Eidelman}},
  \bibinfo {author} {\bibfnamefont {C.}~\bibnamefont {Hanhart}}, \bibinfo
  {author} {\bibfnamefont {A.}~\bibnamefont {Nefediev}}, \bibinfo {author}
  {\bibfnamefont {C.-P.}\ \bibnamefont {Shen}}, \bibinfo {author}
  {\bibfnamefont {C.~E.}\ \bibnamefont {Thomas}}, \bibinfo {author}
  {\bibfnamefont {A.}~\bibnamefont {Vairo}}, \ and\ \bibinfo {author}
  {\bibfnamefont {C.-Z.}\ \bibnamefont {Yuan}},\ }\href {\doibase
  10.1016/j.physrep.2020.05.001} {\bibfield  {journal} {\bibinfo  {journal}
  {Phys. Rept.}\ }\textbf {\bibinfo {volume} {873}},\ \bibinfo {pages} {1}
  (\bibinfo {year} {2020})},\ \Eprint {http://arxiv.org/abs/1907.07583}
  {arXiv:1907.07583 [hep-ex]} \BibitemShut {NoStop}%
\bibitem [{\citenamefont {Chen}\ \emph
  {et~al.}(2021{\natexlab{c}})\citenamefont {Chen}, \citenamefont {Li},
  \citenamefont {Qian}, \citenamefont {Xie}, \citenamefont {Yang},
  \citenamefont {Zhang},\ and\ \citenamefont {Zhang}}]{Chen:2021ftn}%
  \BibitemOpen
  \bibfield  {author} {\bibinfo {author} {\bibfnamefont {S.}~\bibnamefont
  {Chen}}, \bibinfo {author} {\bibfnamefont {Y.}~\bibnamefont {Li}}, \bibinfo
  {author} {\bibfnamefont {W.}~\bibnamefont {Qian}}, \bibinfo {author}
  {\bibfnamefont {Y.}~\bibnamefont {Xie}}, \bibinfo {author} {\bibfnamefont
  {Z.}~\bibnamefont {Yang}}, \bibinfo {author} {\bibfnamefont {L.}~\bibnamefont
  {Zhang}}, \ and\ \bibinfo {author} {\bibfnamefont {Y.}~\bibnamefont
  {Zhang}},\ }\href@noop {} {\  (\bibinfo {year} {2021}{\natexlab{c}})},\
  \Eprint {http://arxiv.org/abs/2111.14360} {arXiv:2111.14360 [hep-ex]}
  \BibitemShut {NoStop}%
\bibitem [{\citenamefont {Chen}\ \emph {et~al.}(2022)\citenamefont {Chen},
  \citenamefont {Chen}, \citenamefont {Liu}, \citenamefont {Liu},\ and\
  \citenamefont {Zhu}}]{Chen:2022asf}%
  \BibitemOpen
  \bibfield  {author} {\bibinfo {author} {\bibfnamefont {H.-X.}\ \bibnamefont
  {Chen}}, \bibinfo {author} {\bibfnamefont {W.}~\bibnamefont {Chen}}, \bibinfo
  {author} {\bibfnamefont {X.}~\bibnamefont {Liu}}, \bibinfo {author}
  {\bibfnamefont {Y.-R.}\ \bibnamefont {Liu}}, \ and\ \bibinfo {author}
  {\bibfnamefont {S.-L.}\ \bibnamefont {Zhu}},\ }\href@noop {} {\  (\bibinfo
  {year} {2022})},\ \Eprint {http://arxiv.org/abs/2204.02649} {arXiv:2204.02649
  [hep-ph]} \BibitemShut {NoStop}%
\bibitem [{\citenamefont {Meng}\ \emph {et~al.}(2022)\citenamefont {Meng},
  \citenamefont {Wang}, \citenamefont {Wang},\ and\ \citenamefont
  {Zhu}}]{Meng:2022ozq}%
  \BibitemOpen
  \bibfield  {author} {\bibinfo {author} {\bibfnamefont {L.}~\bibnamefont
  {Meng}}, \bibinfo {author} {\bibfnamefont {B.}~\bibnamefont {Wang}}, \bibinfo
  {author} {\bibfnamefont {G.-J.}\ \bibnamefont {Wang}}, \ and\ \bibinfo
  {author} {\bibfnamefont {S.-L.}\ \bibnamefont {Zhu}},\ }\href@noop {} {\
  (\bibinfo {year} {2022})},\ \Eprint {http://arxiv.org/abs/2204.08716}
  {arXiv:2204.08716 [hep-ph]} \BibitemShut {NoStop}%
\bibitem [{\citenamefont {Weinberg}(1990)}]{Weinberg:1990rz}%
  \BibitemOpen
  \bibfield  {author} {\bibinfo {author} {\bibfnamefont {S.}~\bibnamefont
  {Weinberg}},\ }\href {\doibase 10.1016/0370-2693(90)90938-3} {\bibfield
  {journal} {\bibinfo  {journal} {Phys. Lett. B}\ }\textbf {\bibinfo {volume}
  {251}},\ \bibinfo {pages} {288} (\bibinfo {year} {1990})}\BibitemShut
  {NoStop}%
\bibitem [{\citenamefont {Weinberg}(1991)}]{Weinberg:1991um}%
  \BibitemOpen
  \bibfield  {author} {\bibinfo {author} {\bibfnamefont {S.}~\bibnamefont
  {Weinberg}},\ }\href {\doibase 10.1016/0550-3213(91)90231-L} {\bibfield
  {journal} {\bibinfo  {journal} {Nucl. Phys. B}\ }\textbf {\bibinfo {volume}
  {363}},\ \bibinfo {pages} {3} (\bibinfo {year} {1991})}\BibitemShut {NoStop}%
\bibitem [{\citenamefont {Bernard}\ \emph {et~al.}(1995)\citenamefont
  {Bernard}, \citenamefont {Kaiser},\ and\ \citenamefont
  {Meissner}}]{Bernard:1995dp}%
  \BibitemOpen
  \bibfield  {author} {\bibinfo {author} {\bibfnamefont {V.}~\bibnamefont
  {Bernard}}, \bibinfo {author} {\bibfnamefont {N.}~\bibnamefont {Kaiser}}, \
  and\ \bibinfo {author} {\bibfnamefont {U.-G.}\ \bibnamefont {Meissner}},\
  }\href {\doibase 10.1142/S0218301395000092} {\bibfield  {journal} {\bibinfo
  {journal} {Int. J. Mod. Phys. E}\ }\textbf {\bibinfo {volume} {4}},\ \bibinfo
  {pages} {193} (\bibinfo {year} {1995})},\ \Eprint
  {http://arxiv.org/abs/hep-ph/9501384} {arXiv:hep-ph/9501384} \BibitemShut
  {NoStop}%
\bibitem [{\citenamefont {Epelbaum}\ \emph {et~al.}(2009)\citenamefont
  {Epelbaum}, \citenamefont {Hammer},\ and\ \citenamefont
  {Meissner}}]{Epelbaum:2008ga}%
  \BibitemOpen
  \bibfield  {author} {\bibinfo {author} {\bibfnamefont {E.}~\bibnamefont
  {Epelbaum}}, \bibinfo {author} {\bibfnamefont {H.-W.}\ \bibnamefont
  {Hammer}}, \ and\ \bibinfo {author} {\bibfnamefont {U.-G.}\ \bibnamefont
  {Meissner}},\ }\href {\doibase 10.1103/RevModPhys.81.1773} {\bibfield
  {journal} {\bibinfo  {journal} {Rev. Mod. Phys.}\ }\textbf {\bibinfo {volume}
  {81}},\ \bibinfo {pages} {1773} (\bibinfo {year} {2009})},\ \Eprint
  {http://arxiv.org/abs/0811.1338} {arXiv:0811.1338 [nucl-th]} \BibitemShut
  {NoStop}%
\bibitem [{\citenamefont {Machleidt}\ and\ \citenamefont
  {Entem}(2011)}]{Machleidt:2011zz}%
  \BibitemOpen
  \bibfield  {author} {\bibinfo {author} {\bibfnamefont {R.}~\bibnamefont
  {Machleidt}}\ and\ \bibinfo {author} {\bibfnamefont {D.~R.}\ \bibnamefont
  {Entem}},\ }\href {\doibase 10.1016/j.physrep.2011.02.001} {\bibfield
  {journal} {\bibinfo  {journal} {Phys. Rept.}\ }\textbf {\bibinfo {volume}
  {503}},\ \bibinfo {pages} {1} (\bibinfo {year} {2011})},\ \Eprint
  {http://arxiv.org/abs/1105.2919} {arXiv:1105.2919 [nucl-th]} \BibitemShut
  {NoStop}%
\bibitem [{\citenamefont {Mei\ss{}ner}(2016)}]{Meissner:2015wva}%
  \BibitemOpen
  \bibfield  {author} {\bibinfo {author} {\bibfnamefont {U.-G.}\ \bibnamefont
  {Mei\ss{}ner}},\ }\href {\doibase 10.1088/0031-8949/91/3/033005} {\bibfield
  {journal} {\bibinfo  {journal} {Phys. Scripta}\ }\textbf {\bibinfo {volume}
  {91}},\ \bibinfo {pages} {033005} (\bibinfo {year} {2016})},\ \Eprint
  {http://arxiv.org/abs/1510.03230} {arXiv:1510.03230 [nucl-th]} \BibitemShut
  {NoStop}%
\bibitem [{\citenamefont {Hammer}\ \emph {et~al.}(2020)\citenamefont {Hammer},
  \citenamefont {K\"onig},\ and\ \citenamefont {van Kolck}}]{Hammer:2019poc}%
  \BibitemOpen
  \bibfield  {author} {\bibinfo {author} {\bibfnamefont {H.~W.}\ \bibnamefont
  {Hammer}}, \bibinfo {author} {\bibfnamefont {S.}~\bibnamefont {K\"onig}}, \
  and\ \bibinfo {author} {\bibfnamefont {U.}~\bibnamefont {van Kolck}},\ }\href
  {\doibase 10.1103/RevModPhys.92.025004} {\bibfield  {journal} {\bibinfo
  {journal} {Rev. Mod. Phys.}\ }\textbf {\bibinfo {volume} {92}},\ \bibinfo
  {pages} {025004} (\bibinfo {year} {2020})},\ \Eprint
  {http://arxiv.org/abs/1906.12122} {arXiv:1906.12122 [nucl-th]} \BibitemShut
  {NoStop}%
\bibitem [{\citenamefont {Rodriguez~Entem}\ \emph {et~al.}(2020)\citenamefont
  {Rodriguez~Entem}, \citenamefont {Machleidt},\ and\ \citenamefont
  {Nosyk}}]{RodriguezEntem:2020jgp}%
  \BibitemOpen
  \bibfield  {author} {\bibinfo {author} {\bibfnamefont {D.}~\bibnamefont
  {Rodriguez~Entem}}, \bibinfo {author} {\bibfnamefont {R.}~\bibnamefont
  {Machleidt}}, \ and\ \bibinfo {author} {\bibfnamefont {Y.}~\bibnamefont
  {Nosyk}},\ }\href {\doibase 10.3389/fphy.2020.00057} {\bibfield  {journal}
  {\bibinfo  {journal} {Front. in Phys.}\ }\textbf {\bibinfo {volume} {8}},\
  \bibinfo {pages} {57} (\bibinfo {year} {2020})}\BibitemShut {NoStop}%
\bibitem [{\citenamefont {AlFiky}\ \emph {et~al.}(2006)\citenamefont {AlFiky},
  \citenamefont {Gabbiani},\ and\ \citenamefont {Petrov}}]{AlFiky:2005jd}%
  \BibitemOpen
  \bibfield  {author} {\bibinfo {author} {\bibfnamefont {M.~T.}\ \bibnamefont
  {AlFiky}}, \bibinfo {author} {\bibfnamefont {F.}~\bibnamefont {Gabbiani}}, \
  and\ \bibinfo {author} {\bibfnamefont {A.~A.}\ \bibnamefont {Petrov}},\
  }\href {\doibase 10.1016/j.physletb.2006.07.069} {\bibfield  {journal}
  {\bibinfo  {journal} {Phys. Lett. B}\ }\textbf {\bibinfo {volume} {640}},\
  \bibinfo {pages} {238} (\bibinfo {year} {2006})},\ \Eprint
  {http://arxiv.org/abs/hep-ph/0506141} {arXiv:hep-ph/0506141} \BibitemShut
  {NoStop}%
\bibitem [{\citenamefont {Fleming}\ \emph {et~al.}(2007)\citenamefont
  {Fleming}, \citenamefont {Kusunoki}, \citenamefont {Mehen},\ and\
  \citenamefont {van Kolck}}]{Fleming:2007rp}%
  \BibitemOpen
  \bibfield  {author} {\bibinfo {author} {\bibfnamefont {S.}~\bibnamefont
  {Fleming}}, \bibinfo {author} {\bibfnamefont {M.}~\bibnamefont {Kusunoki}},
  \bibinfo {author} {\bibfnamefont {T.}~\bibnamefont {Mehen}}, \ and\ \bibinfo
  {author} {\bibfnamefont {U.}~\bibnamefont {van Kolck}},\ }\href {\doibase
  10.1103/PhysRevD.76.034006} {\bibfield  {journal} {\bibinfo  {journal} {Phys.
  Rev. D}\ }\textbf {\bibinfo {volume} {76}},\ \bibinfo {pages} {034006}
  (\bibinfo {year} {2007})},\ \Eprint {http://arxiv.org/abs/hep-ph/0703168}
  {arXiv:hep-ph/0703168} \BibitemShut {NoStop}%
\bibitem [{\citenamefont {Baru}\ \emph {et~al.}(2011)\citenamefont {Baru},
  \citenamefont {Filin}, \citenamefont {Hanhart}, \citenamefont {Kalashnikova},
  \citenamefont {Kudryavtsev},\ and\ \citenamefont {Nefediev}}]{Baru:2011rs}%
  \BibitemOpen
  \bibfield  {author} {\bibinfo {author} {\bibfnamefont {V.}~\bibnamefont
  {Baru}}, \bibinfo {author} {\bibfnamefont {A.~A.}\ \bibnamefont {Filin}},
  \bibinfo {author} {\bibfnamefont {C.}~\bibnamefont {Hanhart}}, \bibinfo
  {author} {\bibfnamefont {Y.~S.}\ \bibnamefont {Kalashnikova}}, \bibinfo
  {author} {\bibfnamefont {A.~E.}\ \bibnamefont {Kudryavtsev}}, \ and\ \bibinfo
  {author} {\bibfnamefont {A.~V.}\ \bibnamefont {Nefediev}},\ }\href {\doibase
  10.1103/PhysRevD.84.074029} {\bibfield  {journal} {\bibinfo  {journal} {Phys.
  Rev. D}\ }\textbf {\bibinfo {volume} {84}},\ \bibinfo {pages} {074029}
  (\bibinfo {year} {2011})},\ \Eprint {http://arxiv.org/abs/1108.5644}
  {arXiv:1108.5644 [hep-ph]} \BibitemShut {NoStop}%
\bibitem [{\citenamefont {Valderrama}(2012)}]{Valderrama:2012jv}%
  \BibitemOpen
  \bibfield  {author} {\bibinfo {author} {\bibfnamefont {M.~P.}\ \bibnamefont
  {Valderrama}},\ }\href {\doibase 10.1103/PhysRevD.85.114037} {\bibfield
  {journal} {\bibinfo  {journal} {Phys. Rev. D}\ }\textbf {\bibinfo {volume}
  {85}},\ \bibinfo {pages} {114037} (\bibinfo {year} {2012})},\ \Eprint
  {http://arxiv.org/abs/1204.2400} {arXiv:1204.2400 [hep-ph]} \BibitemShut
  {NoStop}%
\bibitem [{\citenamefont {Braaten}(2015)}]{Braaten:2015tga}%
  \BibitemOpen
  \bibfield  {author} {\bibinfo {author} {\bibfnamefont {E.}~\bibnamefont
  {Braaten}},\ }\href {\doibase 10.1103/PhysRevD.91.114007} {\bibfield
  {journal} {\bibinfo  {journal} {Phys. Rev. D}\ }\textbf {\bibinfo {volume}
  {91}},\ \bibinfo {pages} {114007} (\bibinfo {year} {2015})},\ \Eprint
  {http://arxiv.org/abs/1503.04791} {arXiv:1503.04791 [hep-ph]} \BibitemShut
  {NoStop}%
\bibitem [{\citenamefont {Schmidt}\ \emph {et~al.}(2018)\citenamefont
  {Schmidt}, \citenamefont {Jansen},\ and\ \citenamefont
  {Hammer}}]{Schmidt:2018vvl}%
  \BibitemOpen
  \bibfield  {author} {\bibinfo {author} {\bibfnamefont {M.}~\bibnamefont
  {Schmidt}}, \bibinfo {author} {\bibfnamefont {M.}~\bibnamefont {Jansen}}, \
  and\ \bibinfo {author} {\bibfnamefont {H.~W.}\ \bibnamefont {Hammer}},\
  }\href {\doibase 10.1103/PhysRevD.98.014032} {\bibfield  {journal} {\bibinfo
  {journal} {Phys. Rev. D}\ }\textbf {\bibinfo {volume} {98}},\ \bibinfo
  {pages} {014032} (\bibinfo {year} {2018})},\ \Eprint
  {http://arxiv.org/abs/1804.00375} {arXiv:1804.00375 [hep-ph]} \BibitemShut
  {NoStop}%
\bibitem [{\citenamefont {Wang}\ \emph
  {et~al.}(2019{\natexlab{a}})\citenamefont {Wang}, \citenamefont {Liu},\ and\
  \citenamefont {Liu}}]{Wang:2018atz}%
  \BibitemOpen
  \bibfield  {author} {\bibinfo {author} {\bibfnamefont {B.}~\bibnamefont
  {Wang}}, \bibinfo {author} {\bibfnamefont {Z.-W.}\ \bibnamefont {Liu}}, \
  and\ \bibinfo {author} {\bibfnamefont {X.}~\bibnamefont {Liu}},\ }\href
  {\doibase 10.1103/PhysRevD.99.036007} {\bibfield  {journal} {\bibinfo
  {journal} {Phys. Rev. D}\ }\textbf {\bibinfo {volume} {99}},\ \bibinfo
  {pages} {036007} (\bibinfo {year} {2019}{\natexlab{a}})},\ \Eprint
  {http://arxiv.org/abs/1812.04457} {arXiv:1812.04457 [hep-ph]} \BibitemShut
  {NoStop}%
\bibitem [{\citenamefont {Meng}\ \emph {et~al.}(2019)\citenamefont {Meng},
  \citenamefont {Wang}, \citenamefont {Wang},\ and\ \citenamefont
  {Zhu}}]{Meng:2019ilv}%
  \BibitemOpen
  \bibfield  {author} {\bibinfo {author} {\bibfnamefont {L.}~\bibnamefont
  {Meng}}, \bibinfo {author} {\bibfnamefont {B.}~\bibnamefont {Wang}}, \bibinfo
  {author} {\bibfnamefont {G.-J.}\ \bibnamefont {Wang}}, \ and\ \bibinfo
  {author} {\bibfnamefont {S.-L.}\ \bibnamefont {Zhu}},\ }\href {\doibase
  10.1103/PhysRevD.100.014031} {\bibfield  {journal} {\bibinfo  {journal}
  {Phys. Rev. D}\ }\textbf {\bibinfo {volume} {100}},\ \bibinfo {pages}
  {014031} (\bibinfo {year} {2019})},\ \Eprint
  {http://arxiv.org/abs/1905.04113} {arXiv:1905.04113 [hep-ph]} \BibitemShut
  {NoStop}%
\bibitem [{\citenamefont {Wang}\ \emph
  {et~al.}(2019{\natexlab{b}})\citenamefont {Wang}, \citenamefont {Meng},\ and\
  \citenamefont {Zhu}}]{Wang:2019ato}%
  \BibitemOpen
  \bibfield  {author} {\bibinfo {author} {\bibfnamefont {B.}~\bibnamefont
  {Wang}}, \bibinfo {author} {\bibfnamefont {L.}~\bibnamefont {Meng}}, \ and\
  \bibinfo {author} {\bibfnamefont {S.-L.}\ \bibnamefont {Zhu}},\ }\href
  {\doibase 10.1007/JHEP11(2019)108} {\bibfield  {journal} {\bibinfo  {journal}
  {JHEP}\ }\textbf {\bibinfo {volume} {11}},\ \bibinfo {pages} {108} (\bibinfo
  {year} {2019}{\natexlab{b}})},\ \Eprint {http://arxiv.org/abs/1909.13054}
  {arXiv:1909.13054 [hep-ph]} \BibitemShut {NoStop}%
\bibitem [{\citenamefont {Wang}\ \emph {et~al.}(2020)\citenamefont {Wang},
  \citenamefont {Meng},\ and\ \citenamefont {Zhu}}]{Wang:2020dko}%
  \BibitemOpen
  \bibfield  {author} {\bibinfo {author} {\bibfnamefont {B.}~\bibnamefont
  {Wang}}, \bibinfo {author} {\bibfnamefont {L.}~\bibnamefont {Meng}}, \ and\
  \bibinfo {author} {\bibfnamefont {S.-L.}\ \bibnamefont {Zhu}},\ }\href
  {\doibase 10.1103/PhysRevD.102.114019} {\bibfield  {journal} {\bibinfo
  {journal} {Phys. Rev. D}\ }\textbf {\bibinfo {volume} {102}},\ \bibinfo
  {pages} {114019} (\bibinfo {year} {2020})},\ \Eprint
  {http://arxiv.org/abs/2009.01980} {arXiv:2009.01980 [hep-ph]} \BibitemShut
  {NoStop}%
\bibitem [{\citenamefont {Meng}\ \emph {et~al.}(2021)\citenamefont {Meng},
  \citenamefont {Wang},\ and\ \citenamefont {Zhu}}]{Meng:2020cbk}%
  \BibitemOpen
  \bibfield  {author} {\bibinfo {author} {\bibfnamefont {L.}~\bibnamefont
  {Meng}}, \bibinfo {author} {\bibfnamefont {B.}~\bibnamefont {Wang}}, \ and\
  \bibinfo {author} {\bibfnamefont {S.-L.}\ \bibnamefont {Zhu}},\ }\href
  {\doibase 10.1016/j.scib.2021.03.016} {\bibfield  {journal} {\bibinfo
  {journal} {Sci. Bull.}\ }\textbf {\bibinfo {volume} {66}},\ \bibinfo {pages}
  {1413} (\bibinfo {year} {2021})},\ \Eprint {http://arxiv.org/abs/2012.09813}
  {arXiv:2012.09813 [hep-ph]} \BibitemShut {NoStop}%
\bibitem [{\citenamefont {Chen}\ \emph
  {et~al.}(2021{\natexlab{d}})\citenamefont {Chen}, \citenamefont {Wang},\ and\
  \citenamefont {Zhu}}]{Chen:2021htr}%
  \BibitemOpen
  \bibfield  {author} {\bibinfo {author} {\bibfnamefont {K.}~\bibnamefont
  {Chen}}, \bibinfo {author} {\bibfnamefont {B.}~\bibnamefont {Wang}}, \ and\
  \bibinfo {author} {\bibfnamefont {S.-L.}\ \bibnamefont {Zhu}},\ }\href
  {\doibase 10.1103/PhysRevD.103.116017} {\bibfield  {journal} {\bibinfo
  {journal} {Phys. Rev. D}\ }\textbf {\bibinfo {volume} {103}},\ \bibinfo
  {pages} {116017} (\bibinfo {year} {2021}{\natexlab{d}})},\ \Eprint
  {http://arxiv.org/abs/2102.05868} {arXiv:2102.05868 [hep-ph]} \BibitemShut
  {NoStop}%
\bibitem [{\citenamefont {Wise}(1992)}]{Wise:1992hn}%
  \BibitemOpen
  \bibfield  {author} {\bibinfo {author} {\bibfnamefont {M.~B.}\ \bibnamefont
  {Wise}},\ }\href {\doibase 10.1103/PhysRevD.45.R2188} {\bibfield  {journal}
  {\bibinfo  {journal} {Phys. Rev. D}\ }\textbf {\bibinfo {volume} {45}},\
  \bibinfo {pages} {R2188} (\bibinfo {year} {1992})}\BibitemShut {NoStop}%
\bibitem [{\citenamefont {Manohar}\ and\ \citenamefont
  {Wise}(2000)}]{Manohar:2000dt}%
  \BibitemOpen
  \bibfield  {author} {\bibinfo {author} {\bibfnamefont {A.~V.}\ \bibnamefont
  {Manohar}}\ and\ \bibinfo {author} {\bibfnamefont {M.~B.}\ \bibnamefont
  {Wise}},\ }\href@noop {} {\emph {\bibinfo {title} {{Heavy quark physics}}}},\
  Vol.~\bibinfo {volume} {10}\ (\bibinfo {year} {2000})\BibitemShut {NoStop}%
\bibitem [{\citenamefont {Zyla}\ \emph {et~al.}(2020)\citenamefont {Zyla} \emph
  {et~al.}}]{ParticleDataGroup:2020ssz}%
  \BibitemOpen
  \bibfield  {author} {\bibinfo {author} {\bibfnamefont {P.~A.}\ \bibnamefont
  {Zyla}} \emph {et~al.} (\bibinfo {collaboration} {Particle Data Group}),\
  }\href {\doibase 10.1093/ptep/ptaa104} {\bibfield  {journal} {\bibinfo
  {journal} {PTEP}\ }\textbf {\bibinfo {volume} {2020}},\ \bibinfo {pages}
  {083C01} (\bibinfo {year} {2020})}\BibitemShut {NoStop}%
\bibitem [{\citenamefont {Du}\ \emph {et~al.}(2020)\citenamefont {Du},
  \citenamefont {Baru}, \citenamefont {Guo}, \citenamefont {Hanhart},
  \citenamefont {Mei\ss{}ner}, \citenamefont {Oller},\ and\ \citenamefont
  {Wang}}]{Du:2019pij}%
  \BibitemOpen
  \bibfield  {author} {\bibinfo {author} {\bibfnamefont {M.-L.}\ \bibnamefont
  {Du}}, \bibinfo {author} {\bibfnamefont {V.}~\bibnamefont {Baru}}, \bibinfo
  {author} {\bibfnamefont {F.-K.}\ \bibnamefont {Guo}}, \bibinfo {author}
  {\bibfnamefont {C.}~\bibnamefont {Hanhart}}, \bibinfo {author} {\bibfnamefont
  {U.-G.}\ \bibnamefont {Mei\ss{}ner}}, \bibinfo {author} {\bibfnamefont
  {J.~A.}\ \bibnamefont {Oller}}, \ and\ \bibinfo {author} {\bibfnamefont
  {Q.}~\bibnamefont {Wang}},\ }\href {\doibase 10.1103/PhysRevLett.124.072001}
  {\bibfield  {journal} {\bibinfo  {journal} {Phys. Rev. Lett.}\ }\textbf
  {\bibinfo {volume} {124}},\ \bibinfo {pages} {072001} (\bibinfo {year}
  {2020})},\ \Eprint {http://arxiv.org/abs/1910.11846} {arXiv:1910.11846
  [hep-ph]} \BibitemShut {NoStop}%
\bibitem [{\citenamefont {Wang}\ \emph {et~al.}(2018)\citenamefont {Wang},
  \citenamefont {Baru}, \citenamefont {Filin}, \citenamefont {Hanhart},
  \citenamefont {Nefediev},\ and\ \citenamefont {Wynen}}]{Wang:2018jlv}%
  \BibitemOpen
  \bibfield  {author} {\bibinfo {author} {\bibfnamefont {Q.}~\bibnamefont
  {Wang}}, \bibinfo {author} {\bibfnamefont {V.}~\bibnamefont {Baru}}, \bibinfo
  {author} {\bibfnamefont {A.~A.}\ \bibnamefont {Filin}}, \bibinfo {author}
  {\bibfnamefont {C.}~\bibnamefont {Hanhart}}, \bibinfo {author} {\bibfnamefont
  {A.~V.}\ \bibnamefont {Nefediev}}, \ and\ \bibinfo {author} {\bibfnamefont
  {J.~L.}\ \bibnamefont {Wynen}},\ }\href {\doibase 10.1103/PhysRevD.98.074023}
  {\bibfield  {journal} {\bibinfo  {journal} {Phys. Rev. D}\ }\textbf {\bibinfo
  {volume} {98}},\ \bibinfo {pages} {074023} (\bibinfo {year} {2018})},\
  \Eprint {http://arxiv.org/abs/1805.07453} {arXiv:1805.07453 [hep-ph]}
  \BibitemShut {NoStop}%
\bibitem [{\citenamefont {Baru}\ \emph {et~al.}(2019)\citenamefont {Baru},
  \citenamefont {Epelbaum}, \citenamefont {Filin}, \citenamefont {Hanhart},
  \citenamefont {Nefediev},\ and\ \citenamefont {Wang}}]{Baru:2019xnh}%
  \BibitemOpen
  \bibfield  {author} {\bibinfo {author} {\bibfnamefont {V.}~\bibnamefont
  {Baru}}, \bibinfo {author} {\bibfnamefont {E.}~\bibnamefont {Epelbaum}},
  \bibinfo {author} {\bibfnamefont {A.~A.}\ \bibnamefont {Filin}}, \bibinfo
  {author} {\bibfnamefont {C.}~\bibnamefont {Hanhart}}, \bibinfo {author}
  {\bibfnamefont {A.~V.}\ \bibnamefont {Nefediev}}, \ and\ \bibinfo {author}
  {\bibfnamefont {Q.}~\bibnamefont {Wang}},\ }\href {\doibase
  10.1103/PhysRevD.99.094013} {\bibfield  {journal} {\bibinfo  {journal} {Phys.
  Rev. D}\ }\textbf {\bibinfo {volume} {99}},\ \bibinfo {pages} {094013}
  (\bibinfo {year} {2019})},\ \Eprint {http://arxiv.org/abs/1901.10319}
  {arXiv:1901.10319 [hep-ph]} \BibitemShut {NoStop}%
\bibitem [{\citenamefont {Golak}\ \emph {et~al.}(2010)\citenamefont {Golak}
  \emph {et~al.}}]{Golak:2009ri}%
  \BibitemOpen
  \bibfield  {author} {\bibinfo {author} {\bibfnamefont {J.}~\bibnamefont
  {Golak}} \emph {et~al.},\ }\href {\doibase 10.1140/epja/i2009-10903-6}
  {\bibfield  {journal} {\bibinfo  {journal} {Eur. Phys. J. A}\ }\textbf
  {\bibinfo {volume} {43}},\ \bibinfo {pages} {241} (\bibinfo {year} {2010})},\
  \Eprint {http://arxiv.org/abs/0911.4173} {arXiv:0911.4173 [nucl-th]}
  \BibitemShut {NoStop}%
\bibitem [{\citenamefont {Liu}\ \emph {et~al.}(2009)\citenamefont {Liu},
  \citenamefont {Liu},\ and\ \citenamefont {Zhu}}]{Liu:2009uz}%
  \BibitemOpen
  \bibfield  {author} {\bibinfo {author} {\bibfnamefont {Y.-R.}\ \bibnamefont
  {Liu}}, \bibinfo {author} {\bibfnamefont {X.}~\bibnamefont {Liu}}, \ and\
  \bibinfo {author} {\bibfnamefont {S.-L.}\ \bibnamefont {Zhu}},\ }\href
  {\doibase 10.1103/PhysRevD.79.094026} {\bibfield  {journal} {\bibinfo
  {journal} {Phys. Rev. D}\ }\textbf {\bibinfo {volume} {79}},\ \bibinfo
  {pages} {094026} (\bibinfo {year} {2009})},\ \Eprint
  {http://arxiv.org/abs/0904.1770} {arXiv:0904.1770 [hep-ph]} \BibitemShut
  {NoStop}%
\bibitem [{\citenamefont {Back}\ \emph {et~al.}(2018)\citenamefont {Back} \emph
  {et~al.}}]{Back:2017zqt}%
  \BibitemOpen
  \bibfield  {author} {\bibinfo {author} {\bibfnamefont {J.}~\bibnamefont
  {Back}} \emph {et~al.},\ }\href {\doibase 10.1016/j.cpc.2018.04.017}
  {\bibfield  {journal} {\bibinfo  {journal} {Comput. Phys. Commun.}\ }\textbf
  {\bibinfo {volume} {231}},\ \bibinfo {pages} {198} (\bibinfo {year}
  {2018})},\ \Eprint {http://arxiv.org/abs/1711.09854} {arXiv:1711.09854
  [hep-ex]} \BibitemShut {NoStop}%
\bibitem [{\citenamefont {Nogga}\ \emph {et~al.}(2005)\citenamefont {Nogga},
  \citenamefont {Timmermans},\ and\ \citenamefont {van Kolck}}]{Nogga:2005hy}%
  \BibitemOpen
  \bibfield  {author} {\bibinfo {author} {\bibfnamefont {A.}~\bibnamefont
  {Nogga}}, \bibinfo {author} {\bibfnamefont {R.~G.~E.}\ \bibnamefont
  {Timmermans}}, \ and\ \bibinfo {author} {\bibfnamefont {U.}~\bibnamefont {van
  Kolck}},\ }\href {\doibase 10.1103/PhysRevC.72.054006} {\bibfield  {journal}
  {\bibinfo  {journal} {Phys. Rev. C}\ }\textbf {\bibinfo {volume} {72}},\
  \bibinfo {pages} {054006} (\bibinfo {year} {2005})},\ \Eprint
  {http://arxiv.org/abs/nucl-th/0506005} {arXiv:nucl-th/0506005} \BibitemShut
  {NoStop}%
\bibitem [{\citenamefont {Birse}\ \emph {et~al.}(1999)\citenamefont {Birse},
  \citenamefont {McGovern},\ and\ \citenamefont {Richardson}}]{Birse:1998dk}%
  \BibitemOpen
  \bibfield  {author} {\bibinfo {author} {\bibfnamefont {M.~C.}\ \bibnamefont
  {Birse}}, \bibinfo {author} {\bibfnamefont {J.~A.}\ \bibnamefont {McGovern}},
  \ and\ \bibinfo {author} {\bibfnamefont {K.~G.}\ \bibnamefont {Richardson}},\
  }\href {\doibase 10.1016/S0370-2693(99)00991-0} {\bibfield  {journal}
  {\bibinfo  {journal} {Phys. Lett. B}\ }\textbf {\bibinfo {volume} {464}},\
  \bibinfo {pages} {169} (\bibinfo {year} {1999})},\ \Eprint
  {http://arxiv.org/abs/hep-ph/9807302} {arXiv:hep-ph/9807302} \BibitemShut
  {NoStop}%
\bibitem [{\citenamefont {Frazer}\ and\ \citenamefont
  {Hendry}(1964)}]{Frazer:1964zz}%
  \BibitemOpen
  \bibfield  {author} {\bibinfo {author} {\bibfnamefont {W.~R.}\ \bibnamefont
  {Frazer}}\ and\ \bibinfo {author} {\bibfnamefont {A.~W.}\ \bibnamefont
  {Hendry}},\ }\href {\doibase 10.1103/PhysRev.134.B1307} {\bibfield  {journal}
  {\bibinfo  {journal} {Phys. Rev.}\ }\textbf {\bibinfo {volume} {134}},\
  \bibinfo {pages} {B1307} (\bibinfo {year} {1964})}\BibitemShut {NoStop}%
\bibitem [{\citenamefont {Eden}\ and\ \citenamefont
  {Taylor}(1964)}]{Eden:1964zz}%
  \BibitemOpen
  \bibfield  {author} {\bibinfo {author} {\bibfnamefont {R.~J.}\ \bibnamefont
  {Eden}}\ and\ \bibinfo {author} {\bibfnamefont {J.~R.}\ \bibnamefont
  {Taylor}},\ }\href {\doibase 10.1103/PhysRev.133.B1575} {\bibfield  {journal}
  {\bibinfo  {journal} {Phys. Rev.}\ }\textbf {\bibinfo {volume} {133}},\
  \bibinfo {pages} {B1575} (\bibinfo {year} {1964})}\BibitemShut {NoStop}%
\bibitem [{\citenamefont {Badalian}\ \emph {et~al.}(1982)\citenamefont
  {Badalian}, \citenamefont {Kok}, \citenamefont {Polikarpov},\ and\
  \citenamefont {Simonov}}]{Badalian:1981xj}%
  \BibitemOpen
  \bibfield  {author} {\bibinfo {author} {\bibfnamefont {A.~M.}\ \bibnamefont
  {Badalian}}, \bibinfo {author} {\bibfnamefont {L.~P.}\ \bibnamefont {Kok}},
  \bibinfo {author} {\bibfnamefont {M.~I.}\ \bibnamefont {Polikarpov}}, \ and\
  \bibinfo {author} {\bibfnamefont {Y.~A.}\ \bibnamefont {Simonov}},\ }\href
  {\doibase 10.1016/0370-1573(82)90014-X} {\bibfield  {journal} {\bibinfo
  {journal} {Phys. Rept.}\ }\textbf {\bibinfo {volume} {82}},\ \bibinfo {pages}
  {31} (\bibinfo {year} {1982})}\BibitemShut {NoStop}%
\bibitem [{\citenamefont {Aaij}\ \emph {et~al.}(2015)\citenamefont {Aaij} \emph
  {et~al.}}]{LHCb:2015yax}%
  \BibitemOpen
  \bibfield  {author} {\bibinfo {author} {\bibfnamefont {R.}~\bibnamefont
  {Aaij}} \emph {et~al.} (\bibinfo {collaboration} {LHCb}),\ }\href {\doibase
  10.1103/PhysRevLett.115.072001} {\bibfield  {journal} {\bibinfo  {journal}
  {Phys. Rev. Lett.}\ }\textbf {\bibinfo {volume} {115}},\ \bibinfo {pages}
  {072001} (\bibinfo {year} {2015})},\ \Eprint
  {http://arxiv.org/abs/1507.03414} {arXiv:1507.03414 [hep-ex]} \BibitemShut
  {NoStop}%
\bibitem [{\citenamefont {Aaij}\ \emph {et~al.}(2019)\citenamefont {Aaij} \emph
  {et~al.}}]{LHCb:2019kea}%
  \BibitemOpen
  \bibfield  {author} {\bibinfo {author} {\bibfnamefont {R.}~\bibnamefont
  {Aaij}} \emph {et~al.} (\bibinfo {collaboration} {LHCb}),\ }\href {\doibase
  10.1103/PhysRevLett.122.222001} {\bibfield  {journal} {\bibinfo  {journal}
  {Phys. Rev. Lett.}\ }\textbf {\bibinfo {volume} {122}},\ \bibinfo {pages}
  {222001} (\bibinfo {year} {2019})},\ \Eprint
  {http://arxiv.org/abs/1904.03947} {arXiv:1904.03947 [hep-ex]} \BibitemShut
  {NoStop}%
\bibitem [{\citenamefont {Molina}\ \emph {et~al.}(2010)\citenamefont {Molina},
  \citenamefont {Branz},\ and\ \citenamefont {Oset}}]{Molina:2010tx}%
  \BibitemOpen
  \bibfield  {author} {\bibinfo {author} {\bibfnamefont {R.}~\bibnamefont
  {Molina}}, \bibinfo {author} {\bibfnamefont {T.}~\bibnamefont {Branz}}, \
  and\ \bibinfo {author} {\bibfnamefont {E.}~\bibnamefont {Oset}},\ }\href
  {\doibase 10.1103/PhysRevD.82.014010} {\bibfield  {journal} {\bibinfo
  {journal} {Phys. Rev. D}\ }\textbf {\bibinfo {volume} {82}},\ \bibinfo
  {pages} {014010} (\bibinfo {year} {2010})},\ \Eprint
  {http://arxiv.org/abs/1005.0335} {arXiv:1005.0335 [hep-ph]} \BibitemShut
  {NoStop}%
\end{thebibliography}%

\end{document}